\documentclass[sigconf]{acmart}
\AtBeginDocument{%
  }

\setcopyright{acmcopyright}
\copyrightyear{2024}
\acmYear{2024}
\acmDOI{XXXXXXX.XXXXXXX}

\acmConference[ASIACCS '24]{ASIACCS '24: In Proceedings of the 2024 ACM Asia Conference on Computer and Communications Security}{July 1--5, 2024}{Singapore}
\acmBooktitle{ASIACCS '24: In Proceedings of the 2024 ACM Asia Conference on Computer and Communications Security, July 1--5, 2024, Singapore}
\acmPrice{15.00}
\acmISBN{978-x-xxxx-XXXX-X/xx/xx}

\usepackage{multirow}

\usepackage{amsmath,amssymb,amsfonts}
\usepackage{algorithmic}
\usepackage{subfigure}
\usepackage{graphicx,xcolor}
\usepackage{textcomp}
\usepackage{booktabs}
\usepackage{enumitem}
\usepackage[framemethod=TikZ]{mdframed}
\usepackage{wasysym}
\usepackage{soul}
\usepackage{pifont}

\newcommand{\us}{$\mu$s}
\newcommand{\mypara}[1]{\vspace{2pt}\noindent\textbf{{#1: }}}
\newcommand{\dramcommand}[1]{\texttt{#1}}
\newcommand{\circled}[1]{\tikz[baseline=(myanchor.base)] \node[circle,fill=.,inner sep=1pt] (myanchor) {\color{-.}\bfseries\footnotesize #1};}

\begin{document}
\title{SoK: Rowhammer on Commodity Operating Systems}
\author{Zhi Zhang}
\email{zzhangphd@gmail.com}
\affiliation{
  \institution{The University of Western Australia}
  \country{Australia}
}

\author{Decheng Chen}
\email{midcchen@mail.scut.edu.cn}
\affiliation{%
  \institution{School of Microelectronics, South China University of Technology}
  \country{China}
}

\author{Jiahao Qi}
\email{qiaoziggy@gmail.com}
\affiliation{%
  \institution{School of Microelectronics, South China University of Technology}
  \country{China}
}

\author{Yueqiang Cheng}
\email{yueqiang.cheng@nio.io}
\affiliation{%
  \institution{NIO}
  \country{China}
}

\author{Shijie Jiang}
\email{jsjscut@gmail.com}
\affiliation{%
  \institution{School of Microelectronics, South China University of Technology}
  \country{China}
}

\author{Yiyang Lin}
\email{milinyiyang@mail.scut.edu.cn}
\affiliation{%
  \institution{School of Microelectronics, South China University of Technology}
  \country{China}
}

\author{Yansong Gao}
\email{gao.yansong@hotmail.com}
\affiliation{%
  \institution{CSIRO's Data61}
  \country{Australia}
}

\author{Surya Nepal}
\email{surya.nepal@data61.csiro.au}
\affiliation{%
  \institution{CSIRO's Data61}
  \country{Australia}
}

\author{Yi Zou}
\email{zouyi@scut.edu.cn}
\affiliation{%
  \institution{School of Microelectronics, South China University of Technology}
  \country{China}
}

\author{Jiliang Zhang}
\email{zhangjiliang@hnu.edu.cn}
\affiliation{%
  \protect\institution{College of Integrated Circuits, Hunan University}
  \country{China}
}

\author{Yang Xiang}
\email{yxiang@swin.edu.au}
\affiliation{%
  \protect\institution{School of Software and Electrical Engineering, Swinburne University of Technology}
  \country{Australia}
}

\renewcommand{\shortauthors}{Zhang et al.}

\begin{abstract}
Rowhammer has drawn much attention from both academia and industry in the past years as rowhammer exploitation poses severe consequences to system security. Since the first comprehensive study of rowhammer in 2014, a number of rowhammer attacks have been demonstrated against dynamic random access memory (DRAM)-based commodity systems to break software confidentiality, integrity and availability. 
Accordingly, numerous software defenses have been proposed to mitigate rowhammer attacks on commodity systems of either legacy (e.g., DDR3) or recent DRAM (e.g., DDR4).
Besides, multiple hardware defenses (e.g., Target Row Refresh) from the industry have been deployed into recent DRAM to eliminate rowhammer, which we categorize as production defenses.

In this paper, we systematize rowhammer attacks and defenses with a focus on DRAM-based commodity systems. Particularly, we have established a unified framework demonstrating how a rowhammer attack affects a commodity system. With the framework, we characterize existing attacks, shedding light on new attack vectors that have not yet been explored. 
We further leverage the framework to categorize software and production defenses, generalize their key defense
strategies and summarize their key limitations, from which potential defense strategies are identified.
\end{abstract}


\maketitle

\section{Introduction}\label{sec:intro}
Recent years have witnessed an infamous hardware vulnerability, termed as \emph{rowhammer}, which is essentially a circuit-level interference problem that exists in 
several mainstream memories (e.g., NAND Flash Memory~\cite{cai2017vulnerabilities}, MRAM~\cite{khan2018analysis} and DRAM~\cite{kim2014flipping}). Among them, rowhammer on DRAM has caused serious security implications as DRAM is widely used in commodity (operating) systems. 

In as early as 2014, Kim et al.~\cite{kim2014flipping} performed the first comprehensive study of rowhammer on DRAM, which showed that frequent accessing (i.e., hammering) DRAM rows (known as \emph{aggressor rows}) could cause bit flips in their adjacent rows (\emph{victim rows}) without accessing the victim rows. In cases where a victim row hosts sensitive data (e.g., page tables), an unprivileged adversary can induce rowhammer-based bit flips to corrupt the data even if legitimate access to the victim row is not allowed, implying that a rowhammer exploit can break memory management unit (MMU)-enforced memory isolation even in the absence of software vulnerabilities, posing severe consequences to the whole system security. Starting from Kim et al.~\cite{kim2014flipping}, a myriad of rowhammer attacks have been demonstrated to break the software's CIA, that is, confidentiality (e.g., leaking sensitive information~\cite{kwong2020rambleed,rakin2021deepsteal,cohen2022hammerscope}), integrity (e.g., gaining privilege escalation~\cite{seaborn2015exploiting,rowhammerjs, van2016drammer, xiao2016one, razavi2016flip, cheng2019cattmew, gruss2017another,frigo2018grand,tatar2018throwhammer,zhang2022implicit}) and availability (e.g., causing denial-of-service to the system~\cite{jang2017sgx,gruss2017another}).

To this end, numerous countermeasures have been proposed and can be classified into two categories, i.e., software and {production} defenses\footnote{While a number of hardware-based defenses (e.g.,~\cite{lee2019twice,park2020graphene,kim2021mithril,yauglikcci2021blockhammer,marazzi2022protrr,juffinger2022csi,saileshwar2022randomized}) have been proposed from the academia, they are beyond our scope as they have not yet been adopted by the industry and are non-applicable to protecting commodity systems.}, among which software solutions %shown in \autoref{fig:rowhammer}, 
have a better deployability by building a defense wall for commodity systems. For production solutions, they were believed to be effective in mitigating the rowhammer attacks until they have been reverse-engineered~\cite{cojocar2019exploiting,frigo2020trrespass,hassan2021uncovering,kogler2022half}.

In the foreseeable future, rowhammer is unlikely to be eliminated and thus continues its threat to system security, making itself continuously heats up the arms race between rowhammer attacks and defenses.
To this end, we are interested in the following questions:

\emph{Can we present a unified framework outlining how a rowhammer attack affects a commodity system from the origin (where it starts) to the end (what it achieves)?
With the framework, can we systematize all the existing works? If so, what are the primary techniques used to mount a rowhammer attack? To mitigate rowhammer attacks, 
by scrutinizing prior works, can we have any novel insights?}

\mypara{Our Work}
To answer the questions above, we perform a systematic study of rowhammer on commodity systems. Specifically, a unified framework is established for mounting an attack, identifying unknown attack vectors and their impacts. Within the framework, we present a taxonomy of existing attacks with a  qualitative analysis of their primary attack techniques. On top of that, we perform an experimental analysis of all the Intel instructions related to CPU cache control and find that \texttt{clwb} can be used for cache flush, which has not discovered by existing attacks. 

Besides, we categorize existing software defenses based on the framework above, from which we can clearly see that all of them aim to mitigate/prevent the attack techniques. However, each category of the defenses has its own limitations in keeping its security guarantee and applicability to commodity systems. 
We further categorize existing {production} defenses based on their security objectives and prototype locations, followed by a summary of their limitations in terms of keeping their security guarantees\footnote{In Section Appendix, we provide a detailed discussion of existing hardware defenses.}.

Last, we provide \emph{explicit research directions} for future attacks and defenses by identifying two potential attacks and three possible defenses. \emph{First}, an attacker in a hardware-assisted VM is likely to trigger rowhammer, break hypervisor-enforced memory isolation and gain hypervisor privilege. \emph{Second}, the attacker may launch an attack from an isolated GPU in x86, particularly in a real-world scenario where GPU and its memory are shared in a multi-tenant public cloud.

As a response, {a defender} may reinforce DRAM-aware memory isolation in hardware-assisted virtualization to address the potential threats from some rowhammer attacks. A practical isolation prototype may reside in a kernel-based virtual machine (KVM), as it is supported by mainstream Linux kernel and widely used by mainstream cloud platforms. 
\emph{Second}, {the defender} may detect rowhammer attacks at a much broader range based on the Intel power-monitoring interface~\cite{lipp2021platypus}, as rowhammer always requires frequent memory accesses to DRAM rows and may generate an abnormal pattern of DRAM power consumption.
\emph{Last}, 
considering that rowhammer requires frequent access to DRAM rows, it will result in an abnormal number of row buffer conflicts. Thus, it is expected to be effective in detecting rowhammer via available Intel uncore performance events.

\mypara{Summary of Contributions}
In summary, we have made three main contributions:

\begin{itemize}[noitemsep, topsep=2pt, partopsep=0pt,leftmargin=0.4cm]
\item We {\it establish} a unified framework showing how a rowhammer attack affects a commodity system from the origin to the end. With the framework, we {\it comprehensively characterize} existing attacks and \emph{identify} possible unknown attack vectors. Particularly, we have analysed primary attack techniques and CPU-cache-control instructions in Intel.
\item We also rely on the unified framework to {\it rigorously categorize} software and {production} defenses, generalize their key defense strategies and summarize their key limitations.
\item We show our insights about {\it concrete research directions} for future rowhammer attacks and defenses.
\end{itemize}
\section{Background and Related Works}\label{sec:background}

\begin{figure}[h]
\centering
\includegraphics[width=\columnwidth]{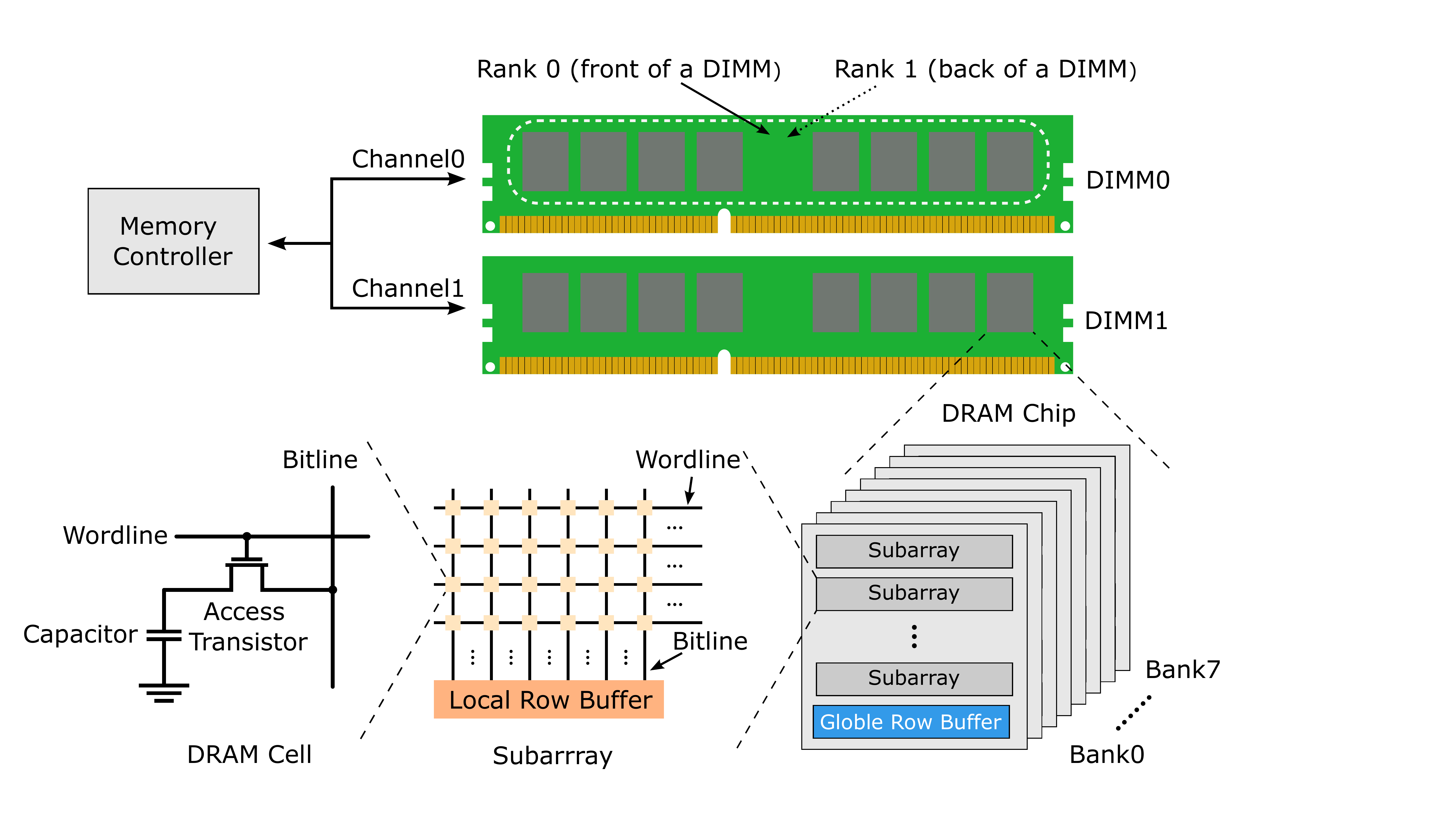}
\caption{DRAM Organization.}
\Description{DRAM Organization.}
\label{fig:mem_org}
\end{figure}

\subsection{DRAM Organization}\label{sec:dram_background}
\autoref{fig:mem_org} presents an overview of a modern DRAM organization.
The memory controller (MC) communicates with dual inline memory modules (DIMMs) through channels and each channel consists of a command bus, an address bus and a data bus. A DIMM has one or more ranks (e.g., single/dual-rank). A rank has a set of chips that operate in lockstep to reply to commands from the MC. Each chip has several banks. {In a typical case where the data bus is 64-bit wide, a rank has 8 chips with 8 banks in a chip to serve the data bus. If multiple ranks share the same  channel, the ranks are multiplexed.} A bank is composed of multiple subarrays and a \emph{global row buffer} (made up of sense amplifiers). Each subarray is a two-dimensional array of cells with a \emph{local row buffer}, storing data from a specific row of cells that is recently accessed. As such, subsequent accesses to the row are
served by the row-buffer. The cells in a row are connected horizontally through a wordline. The cells in a column are connected vertically through a bitline to the local row buffer. The local row buffers are wired to the global row buffer via global bitlines. A cell consists of an access transistor serving as a switch and a capacitor storing a single bit. 

\mypara{DRAM Address Mapping}
{It consists of two main steps.}
In the first {step}, the MC maps physical addresses to logical DRAM addresses. {A {logical} DRAM address refers to a 3-tuple of (bank index, row index, column index), in which the bank index includes DIMM, channel, and rank. }In the second {step}, DRAM internal mapping primarily remaps logical DRAM addresses to physical DRAM addresses. As such, logical rows that are adjacent to each other in the view of the MC are likely to be physically non-adjacent in DRAM.
For the mapping in the first step, it is publicly undocumented but has been reverse-engineered by previous works~\cite{pessl2016drama,xiao2016one,tatar2018defeating,wang2020dramdig} based on a timing side channel~\cite{moscibroda2007memory}.
The remapping in the second step is also kept as confidential and reverse-engineered~\cite{tatar2018defeating,kim2020revisiting,cojocar2020we,orosa2021deeper}.

\subsection{Rowhammer in DRAM}\label{sec:characteristics}
{Prior work has identified characteristics that affect rowhammer-induced bit flips~\cite{kim2014flipping,kim2020revisiting,orosa2021deeper,jiang2021quantifying, Kang2024sledgeHammer} in DRAM-based systems.} In the following, we introduce the major characteristics that are of interest to rowhammer exploits or tests.

\begin{figure}[htbp] 
\centering 
   
    \subfigure[double-sided]{
          \begin{minipage}[t]{0.3\linewidth}
             \centering
             \includegraphics[width=2.5cm, height=2.5cm]{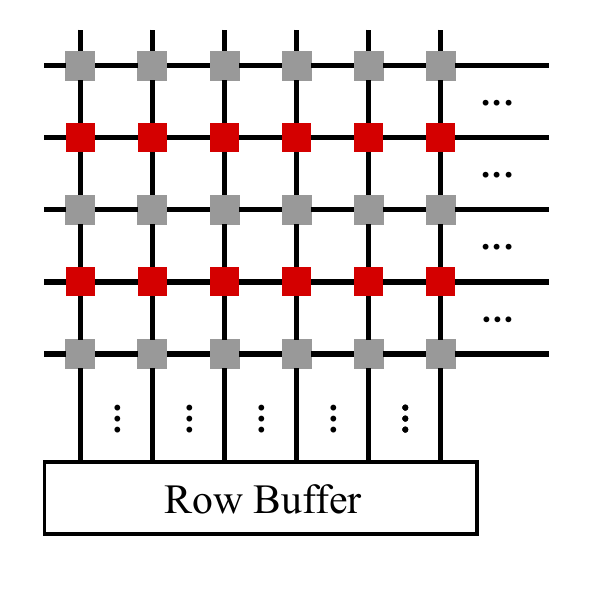}   
             \label{fig:hammer_pattern_a}
             \end{minipage}
    }
    \subfigure[single-sided]{
     \begin{minipage}[t]{0.3\linewidth}
        \centering
            \includegraphics[width=2.5cm, height=2.5cm]{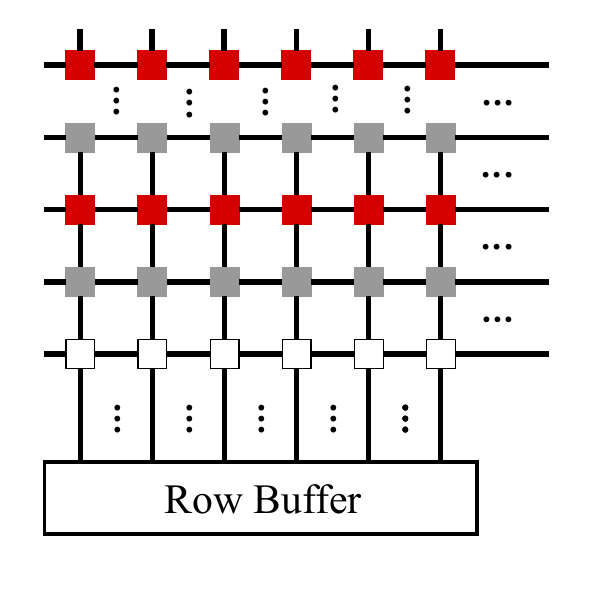} 
            \label{fig:hammer_pattern_b}
         \end{minipage}
    }     
    \subfigure[one-location]{
          \begin{minipage}[t]{0.3\linewidth}
             \centering
             \includegraphics[width=2.5cm, height=2.5cm]{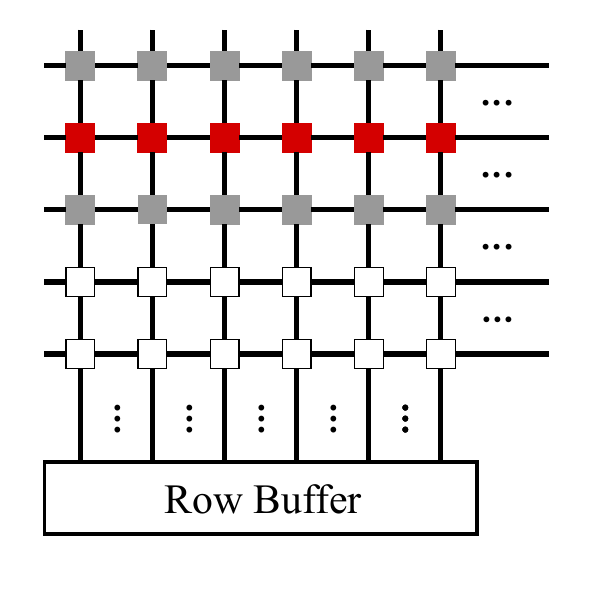}  
             \label{fig:hammer_pattern_c}
             \end{minipage}
    }
    \subfigure[many-sided]{
          \begin{minipage}[t]{0.3\linewidth}
             \centering
             \includegraphics[width=2.5cm, height=2.5cm]{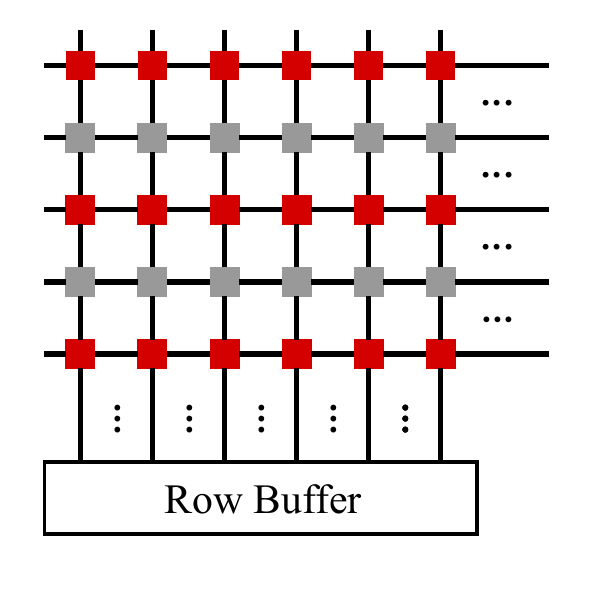} 
             \label{fig:hammer_pattern_d}
             \end{minipage}
    }
    \subfigure[TRR-aided many-sided]{
          \begin{minipage}[t]{0.3\linewidth}
             \centering
             \includegraphics[width=2.5cm, height=2.5cm]{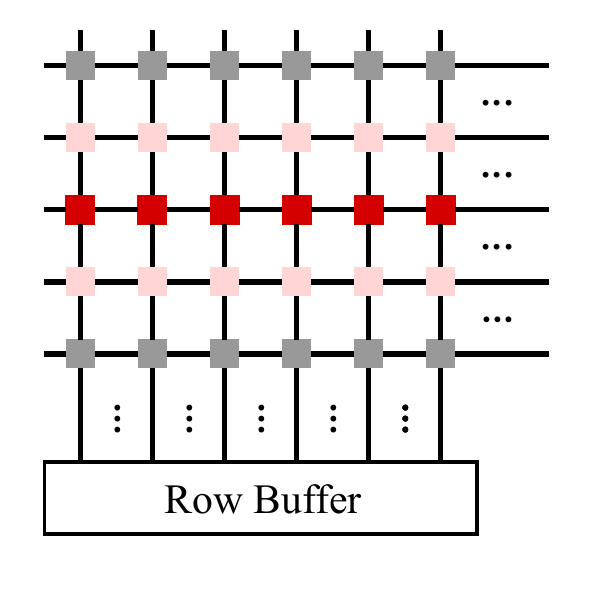} 
             \label{fig:hammer_pattern_e}
             \end{minipage}
    }

    \begin{minipage}[t]{1\linewidth}
        \centering
          \includegraphics[width=8cm,height=0.8cm]{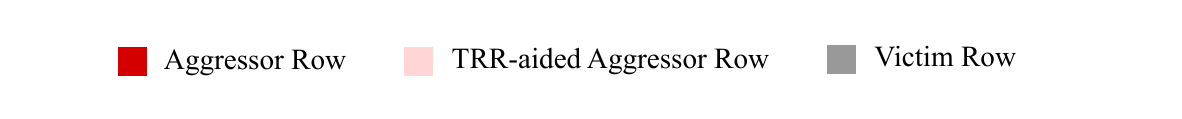}   
    \end{minipage}
            
    \caption{Five existing hammer patterns in the literature. {(We note that rows that are either adjacent or {non}-adjacent to aggressor rows can be vulnerable~\cite{kim2014flipping,kim2020revisiting} and the row distance from an aggressor row {to} a row with bit flips is called \emph{blast radius.})}} 
    \Description{Hammer patterns.}
    \label{fig:hammer_pattern}
\end{figure}

\mypara{Hammer Pattern}
\label{sec:hammer_pattern}
{Kim et al.~\cite{kim2014flipping} found that rowhammer requires repeated accesses (i.e., hammering) to a row to cause permanent charge leakage (i.e., bit flips) within cells of adjacent rows. }As row buffer caches recently accessed rows, it must be bypassed/flushed to ensure memory accesses are directed to rows, i.e., hammering rows. 
A \emph{hammer pattern} refers to how rows are hammered.
As shown in \autoref{fig:hammer_pattern}, there are five existing hammer patterns.  

\emph{Double-sided Hammer} hammers two rows that have one row apart within the same bank. As shown in Figure~\ref{fig:hammer_pattern_a}, all rows adjacent to the hammered rows are likely to be victims, among which the row sandwiched by the two aggressor rows is the most likely to be flipped.
\emph{Single-sided Hammer} randomly selects two rows for hammering, with the hope that the selected rows are within the same bank {and thus the row buffer is bypassed}~\cite{seaborn2015exploiting}. When the rows being hammered happen to be in the same bank, each row neighboring the aggressor row is likely to be a victim. {This is shown in Figure~\ref{fig:hammer_pattern_b}.}  
\emph{One-location Hammer} {is illustrated in Figure~\ref{fig:hammer_pattern_c}} where it only picks one row for hammering~\cite{gruss2017another}. It only applies to specific systems where the MC employs a close-page policy, that is, the MC preemptively closes an accessed row without opening another row, implicitly forcing the DRAM to flush the row buffer. 
\emph{Many-sided Hammer}, {shown in Figure~\ref{fig:hammer_pattern_d}}, hammers more than two rows within the same bank one after another, proposed by TRRespass~\cite{frigo2020trrespass}, resulting in a number of bit flips in DDR4 modules where target row refresh (TRR)~\cite{DDR4TRR} is deployed. 
Following {~\cite{frigo2020trrespass}}, Blacksmith~\cite{jattkeblacksmith} is a variant of this pattern where many aggressor rows are hammered in a non-uniform way, unlike the hammer patterns above where each aggressor row is hammered at the same frequency. 
\emph{TRR-aided Many-sided Hammer} {shown in Figure~\ref{fig:hammer_pattern_e}} abuses TRR to induce bit flips~\cite{kogler2022half}. When an aggressor row with red is hammered, TRR will be triggered to refresh its neighboring rows with pink. Thus, slightly hammering (i.e., a few activations) the pink rows can flip their adjacent victim rows.

\mypara{Hammer Count}
Hammer Count (HC) refers to the number of activations that hammer aggressor rows adjacent to a victim row. {Take the double-sided hammer as an example, the total number of activations to two aggressor rows is referred to as HC~\cite{kim2020revisiting}.} 
DRAM's susceptibility can be quantified by the minimum hammer count to induce the first bit flip, i.e., HC$_{first}$, which can be in the order of 20\,K on DDR3 modules and 10\,K on DDR4 modules~\cite{kim2020revisiting}.

\mypara{Data Pattern}
Values stored in the aggressor and victim rows also significantly affect bit-flip effectiveness, coined as \emph{data pattern}~\cite{kim2014flipping}.
{There are four main data patterns, i.e., Solid (all cells store `0'
or `1'), RowStripe (rows storing `0' are interleaved with
rows storing `1'), ColStripe (columns storing `0'
are interleaved with columns storing `1'), and Checkered (cells storing either `0' or `1' in a checkerboard way). 
Among these patterns, RowStripe {is} the most effective~\cite{kim2014flipping}.}
For a single vulnerable cell, whether it can be flipped depends on not only itself but also the logic values of cells above and below~\cite{kim2014flipping}. For a true cell that is in a charged state and thus stores `1', it can be flipped to `0' most likely when the cells above and below it store `0'{,} but less likely when both cells store `1'. 
If a true cell is in a discharged state and thus stores `0', it cannot be flipped to `1' no matter what the above and below values are. 
For anti cells, their bit-flip direction is also monotonic and works in the opposite way. With this key observation, Ji et al. \cite{ji2019pinpoint} develop effective data patterns at the granularity of a single cell to trigger bit flips in targeted cells while suppressing bit flips in other unwanted cells.

\mypara{Bank-level parallelism}
The aformentioned characteristics focus on hammering a single bank. SledgeHammer~\cite{Kang2024sledgeHammer} can hammer different banks simultaneously with a key observation that while memory accesses to different banks are serialized from the developer's perspective{, they are optimized by hardware to be parallel.} Compared to previous works, it can induce much more bit flips.

\begin{figure*}
\centering
\includegraphics[width=0.9\textwidth]{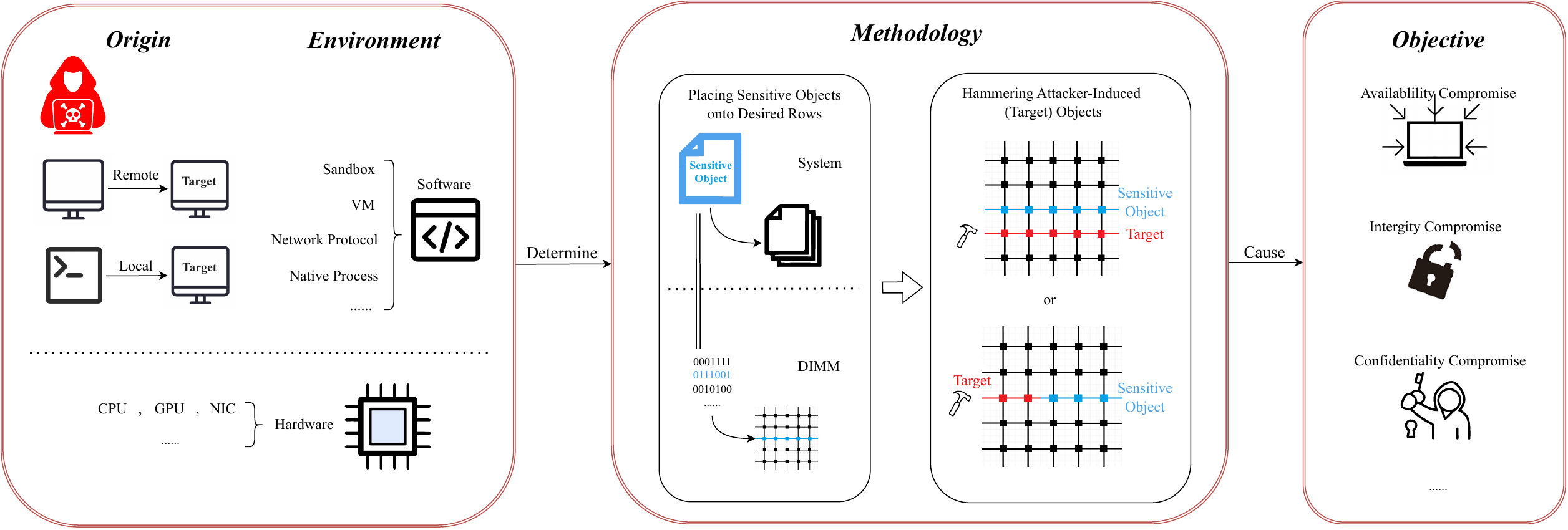}
\caption{A unified framework of rowhammer attacks against commodity systems.}
\Description{A framework of rowhammer attacks}
\label{fig:attack_framework}
\end{figure*}

\subsection{Related Works}
Mutlu et al.~\cite{mutlu2019rowhammer} provided a retrospective of rowhammer in 2019. While they surveyed published rowhammer attacks and defenses at the time of their writing, they discussed each of them succinctly and focused on their previously proposed hardware solution (i.e., PARA~\cite{kim2014flipping}). 
Besides, they discussed the futurespective of rowhammer by describing other potential vulnerabilities besides rowhammer in memory and advocating a principled system-memory co-design approach for rowhammer elimination, which was proposed in their original work~\cite{kim2014flipping} in 2014. 
{In 2021}, Loughlin et al.~\cite{loughlin2021stop} presented a taxonomy of existing mitigations: isolation-centric, frequency-centric and refresh-centric, and proposed new hardware primitives to enable software defenses.

Compared to them, our work has performed a systematization of existing rowhammer attacks and defenses on commodity systems, identified a new instruction for explicit cache flush and described concrete  directions for possible attacks and defenses in real-world. 
\section{Rowhammer Attacks}\label{sec:attacks}
\subsection{Overview of rowhammer attack}\label{sec:overview_attk}
In this section, we present a unified framework, in ~\autoref{fig:attack_framework}, to show how a rowhammer attack affects commodity systems. It consists of four components, which are introduced below.

\mypara{Origin} {Generally, starting a rowhammer attack can be from either local or remote. For local origin, a rowhammer attacker requires crafting their code and executing it in the target system. 
The code is carefully written by different programming languages running on different microarchitectures, resulting in different file formats, e.g., malicious code can be written in Java and C/C++ and compiled into an Android APK~\cite{van2016drammer} on ARM processors, in C/C++ and compiled into an executable binary on x86 processors~\cite{seaborn2015exploiting,gruss2017another,cheng2019cattmew}, and in JavaScript running on either x86 or ARM~\cite{rowhammerjs,frigo2018grand}. Note that the attacker-crafted code execution is performed assuming the attacker resides within the system with an unprivileged access, {e.g., a native process, a browser sandbox, a para-virtualized or hardware-assisted VM, etc. }
In contrast, the attacker from the remote origin \emph{does not} craft code on their own and instead sends network packets only to the target system. It was until~\cite{tatar2018throwhammer, lipp2020nethammer} that network packet transmission was implemented as a viable source of remote origin. }

\mypara{Environment} Starting from the origin, the attacker can reside in various system-defined environments following the principle of least privilege. Thus, the attacker is presented with different attack surfaces and permissions, resulting in different levels of ease to conduct an attack. 
For local origin, the environment is diverse. For example, it can be an unprivileged user process or VM, which allows native binary execution~\cite{xiao2016one,cheng2019cattmew}. In such environments, the attacker is allowed to leverage explicit CPU cache-flush instructions, cache eviction or Direct Memory Access (DMA) to access DRAM directly. For JavaScript code running in a browser sandbox, they have no such permissions and are restricted to cache eviction only. For remote origin-based attacks, they are restricted to DMA. 

\mypara{Methodology} The primary goal of this component is to trigger attacker-desired bit flips, which is achieved through two main steps.

\begin{itemize}[noitemsep, topsep=2pt, partopsep=0pt,leftmargin=0.50cm]
\item[\circled{1}] The attacker aims to place sensitive objects (e.g., cryptographic keys) onto desired rows such that the sensitive objects have close proximity in DRAM to attacker-induced (target) objects, so-called \emph{memory massaging}. 
{The sensitive objects belong to a victim and their integrity or confidentiality can be compromised when placed onto desired rows.} The attacker-induced objects refer to either unprivileged data that are explicitly created and accessed by the attacker, or privileged data that are implicitly created and accessed by the attacker (i.e., a confused deputy). For example, page tables are implicitly created for the attacker's memory mapping and they can be accessed implicitly by the attacker through address translation~\cite{zhang2020pthammer}.

This step can be achieved by abusing various {built-in} features of either software or hardware, resulting in different memory massaging techniques. To evaluate each technique qualitatively, three metrics are defined, i.e., \emph{stealthy} referring to a technique that never causes out-of-memory (OOM), \emph{deterministic} indicating a technique that is definitely able to achieve the aim of this step, and \emph{available} showing a technique that is available by default in the system and has not yet patched by the system.

\item[\circled{2}] The attacker frequently accesses (i.e., hammer) attacker-induced objects in aggressor rows bypassing processor caches and row buffer and achieving desired bit flips. 
As a modern processor has a cache hierarchy to effectively reduce the time of accessing requested data in DRAM memory, access to attacker-induced objects will be first served from the processor caches. 
To enable accessing the objects from the memory, the attacker has to refrain from using the cache of the processor. Alternatively, the attacker can abuse DMA to bypass the cache. 
As row buffer stores recently accessed data in a DRAM bank and acts like the cache, the attacker must also bypass row buffer before each target memory access.  
With bypassing cache and row buffer, the attacker can hammer the attacker-induced objects when accessing them is frequent enough. 

We note that the target objects can be either in the same row with the sensitive objects or adjacent to them as shown in~\autoref{fig:attack_framework}. When they are in the same row, sensitive objects cannot be corrupted. Instead, their information are inferred by flipping bits in their adjacent rows~\cite{kwong2020rambleed}. 

\end{itemize}

\mypara{Objective} The consequences of desired bit flips are devastating to the system security, e.g., {, availability compromise (e.g., denial-of-service), integrity compromise (e.g., privilege escalation) and/or confidentiality compromise (e.g., cryptographic key leakage).} By scrutinizing existing rowhammer attacks via the aforementioned framework, they are categorized based on their origin, environments, memory massaging techniques, and their objectives,
as shown in \autoref{tab:attacks}. In the following, we discuss in detail about the methodology and objective. 

\subsection{Placing Sensitive Objects onto Desired Rows}
\label{sec:placement_attk}
To implement memory massaging, most existing attacks\footnote{A few attacks (i.e., SGX-Bomb~\cite{jang2017sgx}, HammerScope~\cite{cohen2022hammerscope} and Nethammer~\cite{lipp2020nethammer}) do not need memory massaging, as they only require flipping bits in coarse-grained memory locations rather than surgical locations.} have leveraged different system features to propose different techniques. 
In this section, we discuss the techniques based on the features that have been exploited. We also evaluate each technique using the aforementioned three metrics as shown in \autoref{tab:attacks}.

\begin{table*}[!ht]
\footnotesize
\caption{A taxonomy of rowhammer attacks against commodity systems based on the framework in~\autoref{fig:attack_framework}. 
Particularly, the memory massaging technique used by each attack is evaluated via three metrics, i.e., whether it is stealthy, deterministic and/or available (N/A indicates that a work did not use any memory massaging technique).
For the discussion about future explorations, please refer to ~\autoref{sec:insights_attacks}.}
\begin{tabular}{p{1cm}p{2.1cm}llp{2.5cm}lllp{2.8cm}}
\toprule
\multirow{2}{*}{\textbf{Origin}} & \multicolumn{3}{c}{\textbf{Environment}} & \multirow{2}{*}{\textbf{Real-World Attack}} & \multicolumn{3}{c}{\textbf{Memory Massaging}} & \multirow{2}{*}{\textbf{Objective}}  \\ \cline{2-4} \cline{6-8}

&\hspace{0.5cm}{Software} & \multicolumn{2}{l}{\hspace{0.3cm}{Hardware}}& &Stealthy &Deterministic &Available &\\
\hline

\multirow{2}{*}{Remote}  & \multirow{2}{*}{Network Protocol} & \multirow{2}{*}{x86} & \multirow{1}{*}{NIC w/ RDMA} & \multirow{1}{*}{Throwhammer~\cite{tatar2018throwhammer}} &\ding{51} &\ding{53} &\ding{53} & \multirow{1}{*}{Gain app. priv.} \\

& & & CPU & Nethammer~\cite{lipp2020nethammer} &N/A &N/A &N/A & Cause DoS \\ \hline

\multicolumn{1}{l}{\multirow{18}{*}{Local}} & \multirow{4}{*}{Browser Sandbox} &\multirow{4}{*}{x86} &\multirow{4}{*}{CPU} &Bosman et al.~\cite{bosman2016dedup}  &\ding{51} &\ding{51} &\ding{53} & \multirow{4}{*}{Escape sandbox} \\

 &   &  & & SMASH~\cite{de2021smash} &\ding{53} &\ding{51} &\ding{53} &\\

 &   &  & & Seaborn et al.~\cite{seaborn2015exploiting}  &\ding{51} &\ding{53} &\ding{53} &  \\
 
 &   &  &  & Rowhammer.js~\cite{gruss2016rowhammer}  & \ding{53} & \ding{53} &\ding{53} &  \\ \cline{2-9}

 & \multirow{14}{*}{Native Process} &\multirow{14}{*}{x86} &\multirow{14}{*}{CPU} & Seaborn et al.~\cite{seaborn2015exploiting} &\ding{53} &\ding{53} &\ding{51} & \multirow{3}{*}{Gain kernel priv.} \\

 & & & & CATTmew~\cite{cheng2019cattmew} &\ding{51} &\ding{53} &\ding{51} \\

 &   &  &  & PThammmer~\cite{zhang2020pthammer} &\ding{51} &\ding{53} &\ding{51} \\
\cline{5-9}

 &  &  &  & \multirow{2}{*}{Gruss et al.~\cite{gruss2017another}} &\multirow{2}{*}{\ding{51}} &\multirow{2}{*}{\ding{51}} &\multirow{2}{*}{\ding{51}} & Gain root priv. \\ 

 &  &  &  &  & & & & (Cause DoS) \\ 

 &   &  &  & Bhattacharya et al.~\cite{bhattacharya2016curious} &\ding{51} &\ding{53} &\ding{53} & Leak   crypto info. \\ 

 &   &  &  & RAMBleed~\cite{kwong2020rambleed} &\ding{51} &\ding{53} &\ding{51} & Leak crypto info. \\ 
                     
 &   &  & & SGX-Bomb~\cite{jang2017sgx} &N/A &N/A &N/A & Cause DoS \\

 &   &  & & DeepSteal~\cite{rakin2021deepsteal} &\ding{51} &\ding{53} &\ding{51} & Leak DNN weights \\
 
 &   &  & & SpecHammer (kernel exploit)~\cite{tobahspechammer} &\ding{53} &\ding{53} &\ding{51} & Strengthen Spectre \\

  &   &  & & GadgetHammer~\cite{YTobah2024gogogadget} &\ding{53} &\ding{53} &\ding{51} & Leak kernel info. \\
  
 &   &  & & \multirow{1}{*}{HammerScope~\cite{cohen2022hammerscope}} &N/A &N/A &N/A & Break KASLR \\

 \cline{5-9}  
 &   &  & & Hong et al.~\cite{hong2019terminal} &\ding{51} &\ding{51} &\ding{53} &\multirow{2}{*}{Degrade DNN model} \\
 
 &   &  & & DeepHammer~\cite{yao2020deephammer} &\ding{51} &\ding{53} &\ding{51} \\
 
 \cline{2-9}

 &ParaVM &x86 &CPU & Xiao et al.~\cite{xiao2016one} &\ding{51} &\ding{51} &\ding{51} & Gain hypervisor priv. \\ \cline{2-9}

&{\multirow{2}{*}{HVM}} &\multirow{2}{*}{x86}  &\multirow{2}{*}{CPU}  & Razavi et al.~\cite{razavi2016flip}  &\ding{51} &\ding{51} &\ding{53} & Gain HVM priv. \\
\cline{5-9}
& & & &\textbf{Future Exploration} & & & &{Gain hypervisor priv.} \\ 
\cline{1-9}
 &  &x86 &{\textbf{Isolated GPU}} & \multicolumn{4}{c}{\textbf{Future Exploration}} \\
\midrule

{\multirow{2}{*}{Local}} & Browser Sandbox & {ARM} & Integrated GPU & GLitch~\cite{frigo2018grand} &\ding{51} &\ding{51} &\ding{51} & Escape sandbox \\\cline{2-9}

& \multirow{2}{*}{Native Process}  &\multirow{2}{*}{ARM} &\multirow{2}{*}{CPU}  & Drammer~\cite{van2016drammer} &\ding{53} &\ding{51} &\ding{53} & \multirow{2}{*}{Gain kernel priv.} \\
                     
&  &  &  & RAMpage~\cite{van2018guardion} &\ding{53} &\ding{51} &\ding{53} &  \\
\bottomrule
\end{tabular} 

\label{tab:attacks}
\end{table*}
\mypara{\texttt{mmap}}
It is a POSIX user interface, by which a user can specify a file to be memory-mapped and access it from her own address space. {In the case of Linux}, the kernel loads the file content into physical memory and allocates page-table entries accordingly in an on-demand {manner}, that is, the physical memory and page-table entries are not allocated until the file is accessed by the user process. 
Thus, the attacker can repeatedly invoke this feature, forcing the kernel to create a large number of page-table pages. These pages are sensitive objects and some {are} highly likely to be physically adjacent to the attacker pages in aggressor rows~\cite{seaborn2015exploiting,cheng2019cattmew,zhang2020pthammer}. {However, if \texttt{mmap} is abused excessively, available memory will be exhausted, resulting in OOM~\cite{seaborn2015exploiting}.}

\mypara{\texttt{fork}}
It is a system call, by which a (parent) process can create a new child process. Similar to the invoking process, the child process has its own kernel structures, such as \texttt{struct task\_struct} and \texttt{struct cred}. Among the structures, \texttt{struct cred} is {of interest to an attacker}, as it stores the user id of a process. 
{An attacker can} repeatedly invoke \texttt{fork} to create a large number of processes and thus sprays the memory with sensitive structures, with the hope that some \texttt{struct cred} will be placed next to the attacker pages in aggressor rows~\cite{cheng2019cattmew,zhang2020pthammer}.

\mypara{WebGL} 
It is a JavaScript API for web developers to accelerate 2D and 3D graphics rendering {in} mainstream browsers. WebGL reserves a memory cache pool for {processing textures in integrated GPUs}, and the pool contains {2048} pages. To prevent vulnerable pages of textures from being freed to the pool, GLitch~\cite{frigo2018grand} releases 2048 previously-allocated textures to the pool. Right after releasing vulnerable textures, it applies for a large memory allocation of target JavaScript objects \texttt{ArrayObjects}. Thus, some \texttt{ArrayObjects} are predicted to reuse the freed vulnerable texture pages and thus corrupted by rowhammer.

\mypara{Page deduplication}
To improve memory utilization, a running system merges physical pages that have the same content, coined as page deduplication. This feature allows {cloud service providers} to run more VMs with the same {amount of} physical memory. 
To abuse this feature, the attacker crafts an attacker page at a specified vulnerable page. This page has the same content {as the} target sensitive page of another {VM} in a cloud setting. Thus, the underlying KVM hypervisor is lured into merging the two pages into the attacker-specified page~\cite{razavi2016flip}. Similarly, Bosman et al.~\cite{bosman2016dedup} abused this feature for rowhammer attacks against Microsoft Edge browser. However, this feature has been disabled in commodity systems~\cite{win10,redhat}, defeating rowhammer attacks exploiting this feature~\cite{razavi2016flip,bosman2016dedup,hong2019terminal}.

\mypara{Per-CPU page-frame cache update policy}
Linux buddy allocator divides system memory into multiple non-overlapping \emph{zones}, e.g., \emph{ZONE\_DMA}, \emph{ZONE\_DMA32} and \emph{ZONE\_normal} in the x86-64 architecture. 
If a process running on a CPU frees a page frame, the buddy allocator does not return the freed page frame to its corresponding zone. Instead, it pushes the freed page frame into a per-CPU page-frame cache (i.e., \texttt{per-CPU pageset}), which maintains a list of free page frames released by the CPU using the policy of Last-In-First-Out (LIFO). A consequence of this design is that the freed page frame is likely to be reused by the CPU and the page content is still hot in the per-CPU cache. 
RAMBleed~\cite{kwong2020rambleed}, DeepHammer~\cite{yao2020deephammer} and SpecHammer (user exploit)~\cite{tobahspechammer} first hold vulnerable page frames. Then they free the page frames and trigger targeted processes to run. With the update policy, the page frame cache will allocate the freed page frames for targeted sensitive pages of the processes.

\mypara{Page frame allocation policy}
Linux buddy allocator manages system memory as memory blocks and each block has a continuous physical memory region, the size of which is a power-of-two number of page frames. For blocks that have the same size, they are maintained in the same block list. 
To serve a memory allocation request, Linux first searches each block list iteratively for a block that satisfies the request. If none of the blocks have the requested size, Linux splits a larger-sized block into two blocks, returns one block to fulfill the request, and adds the other block into an appropriate block list. 
Upon a memory deallocation request, Linux tries to merge the freed page frames with their neighboring free page frames (if possible),  generating a bigger block and updating it into a block list. This allocation policy can minimize external fragmentation of physical memory.

Drammer~\cite{van2016drammer} and RAMpage~\cite{van2018guardion} abuse this allocation policy by exhausting and freeing memory blocks in a predictable way, with the goal of placing a page-table page onto an available and vulnerable page frame. {However, they are highly like to cause OOM.}{ Besides, to abuse the policy, both attacks must use the Android ION interface, which has been deprecated by recent android kernel versions.} SpecHammer (kernel exploit)~\cite{tobahspechammer} and GadgetHammer~\cite{YTobah2024gogogadget} leverage the policy to create memory pressure, which will probably force the kernel to use a vulnerable user page to store targeted kernel stack variables. 
SMASH~\cite{de2021smash} consumes all the memory blocks that are smaller than 2\,MB and then requests additional 4\,KB page frames for JavaScript \texttt{ArrayBuffers}, which  forces the buddy allocator to split a previously freed vulnerable 2\,MB memory block. The split vulnerable 4\,KB page frame is predicted to host JavaScript \texttt{ArrayBuffer}. {If many blocks smaller than than 2\,MB are exhausted, it might cause OOM. Besides, 2\,MB memory blocks are supported by transparent huge pages (THPs), which are not available by default in the recent browsers.} This non-default configuration also renders another browser-oriented exploit (i.e., rowhammer.js~\cite{gruss2016rowhammer}) ineffective.
CATTmew~\cite{cheng2019cattmew} exhausts the memory blocks that are smaller than or equal to the DIMM row size (e.g., 256\,KB). Then the buddy allocator is forced to split a large memory block that is twice the row-size when the attacker allocates 4\,KB attacker pages and uses \texttt{mmap} for 4\,KB page-table page allocations. By doing so, page-table pages are likely to {be} placed onto vulnerable pages next to the attacker pages.

\mypara{memcached item allocation policy}
A memcached architecture manages key-value pairs, which are stored in-memory as \emph{items}. The corresponding data structure for each item is \texttt{struct \_stritem} and it has different sizes ranging from 96\,bytes to 1\,MB. The memcached maintains a singly linked list, i.e., hash chain, which stores different keys but with the same hashes (i.e., colliding keys). The hash is computed by 
a hash function with the key as an input. However, the function is not cryptographically secure and thus its computed hash value can be resulted from different keys. 
First, the attacker~\cite{tatar2018throwhammer}, as a remote client process, issues numerous \texttt{SET} requests to the memcached  for crafting many 1\,MB items in which each key is hashed to the same value. These items are chained together and can be read by the attacker.
Then, the attacker sends a \texttt{DELETE} request to force the memcached to free a target 1\,MB item, and then issues \texttt{GET} requests to lure the memcached into reusing the 1\,MB item for smaller-sized item allocations, with the hope that one smaller item will land on a vulnerable DRAM location.
{However, \texttt{DELETE} is not supported by default in recent versions of RDMA-based memcached~\cite{tatar2018throwhammer}.}

\mypara{Page cache eviction policy}
To boost system performance, Linux manages unused page frames as page cache. If a memory-mapped file is released by a process, Linux still keeps its data in the page cache. If the file is accessed again, Linux does not have to load the file from non{-}volatile storage such as disks and instead serves the access quickly from page cache. 
Linux considers pages in the page cache {as} available memory and thus will evict obsolete file data from these pages to load recently accessed files. 
Gruss et al.~\cite{gruss2017another} observe that if Linux evicts file data from a page-cache page frame, it reloads the evicted file data onto a different page frame upon access. By abusing the eviction policy repeatedly, they relocate sensitive code of a \texttt{setuid} process onto vulnerable pages~\cite{gruss2017another} deterministically.

\mypara{Page table update policy in paravirtualization} {In paravirtualization, guest OS inside each VM is modified to work with its underlying hypervisor.}
In Xen paravirtualization, MMU of a paravirtualized {VM} is modified to fill its page table entries with physical page frame numbers (PFNs) rather than pseudo PFNs. To isolate each VM, Xen hypervisor enforces an invariant that a VM cannot write its page tables directly. Instead, the VM must invoke a hypercall for page table updates. However, the VM is allowed to specify page frames that host its page tables. 
With this key observation, Xiao et al.~\cite{xiao2016one} place {a target} page-table page onto a specified vulnerable page frame without requiring much memory.

\subsection{Hammering Attacker-Induced Objects}
\label{sec:hammer_attk}

\begin{table}
\footnotesize
\centering
\caption{A taxonomy of techniques proposed by rowhammer attacks in avoiding CPU cache, GPU cache and row buffer.}
\label{tab:techniques}
\begin{tabular}{lll}
\toprule
 \multicolumn{2}{c}{\textbf{Objective}} & \multicolumn{1}{l}{\textbf{Technique}} \\ \hline
\multirow{6}{*}{\textbf{Avoid using}} & \multirow{4}{*}{\textbf{CPU cache}} &  explicit cache-flush instructions   \\                    
                                      &       & non-temporal instructions  \\ 
                                      &       & \multicolumn{1}{l}{(Intel CAT-assisted) cache eviction}                           \\ 
                                      &       & \multicolumn{1}{l}{direct memory access}                    \\ \cline{2-3}                                  
                                      &  \multirow{1}{*}{\textbf{GPU cache}}     & \multicolumn{1}{l}{cache eviction}
                                      \\
                                      \cline{2-3}
                                      &\multicolumn{1}{l}{\multirow{1}{*}{\textbf{row buffer}}} & \multicolumn{1}{l}{\multirow{1}{*}{{different hammer patterns}}} \\
                                     
\bottomrule
\end{tabular}
\end{table}

\subsubsection{Refraining from using CPU/GPU cache}
As shown in \autoref{tab:techniques}, there are four techniques that have been proposed to avoid using either CPU or GPU cache and thus induce a DRAM memory access, that is, \emph{explicit cache-flush instruction}, \emph{non-temporal instruction}, \emph{cache eviction} and \emph{direct memory access}. In the following, each technique is introduced in detail. 

\mypara{Explicit cache-flush instruction}
{Modern CPUs have multiple levels of caches. For each level of cache, it consists of cache sets and each cache set consists
of multiple cache lines (also knowns as ways).}
Existing x86-based rowhammer attacks have identified a couple of unprivileged instructions to explicitly flush a CPU cache line specified by a memory address from all levels of cache hierarchy, i.e., \texttt{clflush} and \texttt{clflushopt}. \texttt{clflush} is available in all Intel-based microarchitectures while \texttt{clflushopt} is only available since the recent Intel 6th generation microarchitectures (e.g., SkyLake).

Besides, we perform an analysis of all the Intel instructions relevant to CPU cache control~\cite{intelOp} (i.e., \texttt{clflush}, \texttt{clflushopt}, \texttt{clwb} and \texttt{cldemote}) and observe that \texttt{clwb}, as an unprivileged instruction, can also be used for explicit cache flush. Specifically, 
\texttt{clwb} is used to write a modified cache line (represented by a linear address) back to memory. After that, the cache line may be retained in the cache hierarchy or may be invalidated from the cache hierarchy by hardware (an indicator of cache flush).

Our empirical observation is performed on a Supermicro server with Intel Xeon Gold 6230 CPU (a Cascade Lake microarchitecture)
as \texttt{clwb} is only available in certain Intel microarchitectures (e.g., Intel Xeon Processor Scalable Family and 11th Generation Intel Core processor family). 
Based on Mastik\footnote{Mastik is a microarchitectural side-channel toolkit and available at \newline https://github.com/0xADE1A1DE/Mastik}, we design a finite loop that runs 1,000,000 times, within which \texttt{mfence} {(a memory barrier instruction in Intel x86 used to ensure memory ordering.)} is followed by measuring the access latency to a linear address from an allocated memory buffer. The access latency is then averaged. 
As a comparison, \texttt{mfence} is replaced by an instruction sequence of \texttt{clwb} and \texttt{mfence}. The results show that the access with \texttt{clwb} takes about 245 CPU cycles while the access without \texttt{clwb} costs around 20 CPU cycles, showing that \texttt{clwb} has explicitly flushed a specified cache line.
Further, we measure the elapsed CPU cycles of invoking each explicit cache-flush instruction. Each invocation is also repeated for 1,000,000 times in a finite loop, generating an averaged measurement.  
For \texttt{clwb}, \texttt{clflushopt} and \texttt{clflush}, their respective averaged cycles are 252, 263 and 287, showing that \texttt{clwb} outperforms the other two in their hammering efficiency. 

\mypara{Non-temporal instruction}
Given that unprivileged non-temporal instructions used for non-temporal data references employ cache bypass, the attacker can also invoke non-temporal store instructions to perform write access to data directly into memory without CPU cache involved.
Qiao et al.~\cite{qiao2016new} and BitMine~\cite{zhang2021bitmine} thoroughly analyze non-temporal store instructions on Intel microarchitectures and present a string {of} available instructions, i.e., \texttt{movnti}, \texttt{movntdq},
\texttt{movntpd}, \texttt{movntps}, \texttt{movntq}, \texttt{maskmovq} and \texttt{maskmovdqu}.

\mypara{Cache eviction}
As CPU/GPU cache has limited capacity, the attacker can evict a target address from either CPU~\cite{gruss2016rowhammer} or GPU~\cite{frigo2018grand} by accessing a {set of many enough virtual addresses}. The addresses in the set are congruent in a way that they are mapped to the same {cache set} and the same cache slice as the target address. This technique is particularly useful when launching an attack in a browser sandbox environment where the cache-flush and non-temporal instructions are not available. 
As Intel cache allocation technology (CAT)~\cite{herdrich2016cache} allows system software to partition the CPU cache, each subset of the cache is dedicated to a process or virtual machine to mitigate cache thrashing. 
In such a case, the cache capacity for a process decreases significantly, indicating that the number of congruent addresses in the eviction set will decrease. Aga et al.~\cite{aga2017good} have abused Intel CAT to facilitate rowhammer attacks.

\mypara{Direct memory access}
Last, the attacker can access uncached memory directly by abusing the DMA feature. The DMA memory is marked as uncached and thus accessing it will bypass the CPU cache. Unlike the non-temporal instructions that require write access, DMA access can be any type, i.e., write, read or execute. In the x86 architecture, both Throwhammer~\cite{tatar2018throwhammer} and Nethammer~\cite{lipp2020nethammer} launch the attack by sending network packets directly into DMA memory. In the ARM platform, 
Drammer~\cite{van2016drammer} can access DMA memory even in a local unprivileged process.

\subsubsection{Bypassing Row buffer}\label{sec:bypass_rowbuffer}
As row buffer resides in each bank, the processor and peripheral devices cannot clear row buffer explicitly. To this end, the attacker can leverage the hammer patterns introduced in \autoref{sec:hammer_pattern} to clear row buffer implicitly. In the single-sided hammer, the attacker accesses several randomly picked addresses besides the target-object addresses. Clearly, these two hammer patterns are much less effective than double-sided hammer in DDR3 modules and many-sided hammer in DDR4 modules. To implement double-sided hammer or many-sided hammer, the attacker must know which rows to access and thus requires (partial) mappings from virtual addresses to physical addresses, physical addresses to logical DRAM addresses and logical to physical DRAM address.

For the last two mappings, they are either available from public documentation or have been reverse-engineered as discussed in \autoref{sec:dram_background}.
For the first mapping, an unprivileged process can acquire it through the \emph{pagemap} interface but the interface has been restricted to root users since Linux version 4.0~\cite{shutemovpagemap} to mitigate rowhammer. Alternatively, the attacker can abuse the feature of huge page (e.g., 2\,MB) in the x86 architecture or Android ION in the ARM architecture to request a large block of contiguous physical memory. 
Within the large memory, it is highly likely to find multiple same-bank rows. However, huge page is disabled by default in the browser sandbox environment. To address this issue, SledgeHammer~\cite{Kang2024sledgeHammer} leverages a contention-based cache side channel in verifying the contiguity of memory blocks. By doing so, they are able to find many 2\,MB  blocks that physically contiguous.

\subsubsection{Frequently Accessing Attacker-Induced Objects}
Most attacks require putting objects explicitly created by the attacker onto aggressor rows and applying different techniques in \autoref{tab:techniques} to avoid processor caches and row buffer for hammering the objects, which is coined as \emph{explicit hammer}~\cite{zhang2020pthammer}. 

For objects that are implicitly created, the attacker cannot access them explicitly and instead can use a benign entity to implicitly access the objects, the so-called \emph{implicit hammer}~\cite{zhang2020pthammer}. The entity can be hardware (e.g., the processor) or software (e.g., system call handler). For instance, PThammer uses the page-table-walk feature of CPU while GhostKnight~\cite{zhang2020ghostknight} abuses the speculative execution feature of CPU for implicit hammer.
To trick the benign entity into frequently accessing the objects, the effective way for bypassing the caches is restricted to cache eviction~\cite{zhang2020pthammer}.

\subsection{Objective}
\label{sec:exploit_attk}

\subsubsection{Causing Denial-of-service}
An attacker can demonstrate a DoS attack either locally or remotely.   
Both SGX-Bomb~\cite{jang2017sgx} and Gruss et al.~\cite{gruss2017another} have demonstrated local DoS attack by abusing a feature of Intel SGX~\cite{costan2016intel}, i.e., data integrity check. Specifically, an SGX enclave has a physically contiguous memory region, which is encrypted in DRAM and protected from non-enclave memory accesses. If data in the enclave memory is corrupted, the corruption will be detected by SGX's memory encryption engine (MEE), resulting in a system halt. With this observation, the attacker can simply apply for an SGX enclave and induce a single bit flip in the enclave memory to make the whole system unresponsive.

Compared to local attacks, remote attacks do not require code execution from the attacker. Specifically, Nethammer~\cite{lipp2020nethammer} simply sends crafted network packets repeatedly to a target system. To handle the packets, the system frequently accesses (read, write or execute) relevant memory. If the system applies techniques in \autoref{tab:techniques}, bit flips can be induced remotely in the system and some of them can occur in the file system, thus crashing the entire system.

\subsubsection{Gaining Privilege Escalation}
Various rowhammer attacks break different forms of memory isolation on different microarchitectures with the goal of achieving privilege escalation. These attacks are discussed based on their targeted privilege software. 

\mypara{Targeting browser}
A sandbox provides a browser-controlled environment where an untrusted program is running with restricted permissions, separating untrusted code from trusted code within the same process. 
There are five attacks that have leveraged a malicious program to break the sandbox environment and gain privilege escalation. Among them, one uses native code and four are based on JavaScript code. 

Google native client provides a sandboxing environment to run validated instructions from mainstream architectures such as x86 and ARM. Seaborn et al.~\cite{seaborn2015exploiting} demonstrate a Chrome sandbox escape. Particularly, a malicious native client program uses bit flips to corrupt opcodes of validated instructions and turns {them} into unsafe ones, enabling arbitrary calls of host OS's \texttt{syscalls}.
Gruss et al.~\cite{rowhammerjs} escape a Firefox browser by crafting JavaScript code in a website and corrupting page tables through bit flips. SMASH~\cite{de2021smash} corrupts metadata of a JavaScript \texttt{ArrayBuffer} and compromises the latest Firefox browser of 2021.
Bosman et al.~\cite{bosman2016dedup} compromise a Microsoft Edge browser through rowhammer. They abuse the page deduplication feature to craft counterfeit JavaScript \texttt{Uint8Array} objects and corrupt the objects with bit flips, resulting in an arbitrary read/write primitive within the browser.
GLitch~\cite{frigo2018grand} is an attack against integrated graphics processing unit (GPU) on a mobile platform, which corrupts elements of JavaScript \texttt{ArrayObject} and escapes a Firefox sandbox.

\mypara{Targeting memcached server application}
A memcached server application allows a remote client to send network packets directly into a specified memory region, i.e., remote DMA (RDMA).
Throwhammer~\cite{tatar2018throwhammer} uses RDMA-enabled NIC to compromise the entire memcached server application with arbitrary write and code execution. 
It abuses the memcached item allocation policy to corrupt memcached items by simply issuing network packets without requiring any local code execution.

\mypara{Targeting \texttt{setuid} process}
\texttt{setuid} is for set user ID on execution, which is a type of file permission in Unix-like operating systems such as Linux. When an unprivileged user launches a binary with \texttt{setuid}, the binary will run with root privilege rather than the user privilege. 
Gruss et al.~\cite{gruss2017another} mount an attack from an unprivileged process against a \texttt{setuid} process. They break MMU-enforced inter-process isolation and bypass all existing defenses prior to their proposed attack.
Particularly, they abuse the page cache eviction policy to corrupt opcodes of critical branches in the \texttt{setuid} process, resulting in root privilege escalation. 

\mypara{Targeting Kernel}
{Kernel is one of the most appealing targets.} There are four attacks breaking user-kernel isolation and corrupting page tables, resulting in kernel privilege escalation from an unprivileged native process.

Specifically, Seaborn et al.~\cite{seaborn2015exploiting} mount the first rowhammer attack against kernel in the x86 architecture. They propose abusing the \texttt{mmap} interface to spray the last-level page-table pages, some of which will be placed adjacent to attacker-accessible pages. 
CATTmew~\cite{cheng2019cattmew} demonstrates an attack against CATT~\cite{brasser17can}, the first DRAM-aware user-kernel isolation. 
They have observed the so-called memory-ownership weakness of CATT, that is, there exist kernel buffers that are still accessible to unprivileged users in the presence of CATT. With this observation, they hammer video buffers or SCSI Generic buffers and trigger bit flips in some page tables created by \texttt{mmap}. PThammer~\cite{zhang2020pthammer} introduces implicit hammer-based attack,
which tricks the CPU into hammering page tables, resulting in other page table corruption. Drammer~\cite{van2016drammer} performs an attack against the ARM architecture. It abuses the page frame allocation policy of Linux buddy allocator to surgically induce bit flips in a target page-table entry.

\mypara{Targeting paravirtualized hypervisor}
In Xen paravirtualization, a VM has no write access but only read access to its own page tables. Thus, the VM's page table updates must be approved by the underlying hypervisor. 
Xiao et al.~\cite{xiao2016one} propose an attack against the paravirtualized hypervisor. They abuse the page table update policy in paravirtualization to flip bits in page tables and gain write access to them, breaking the MMU-enforced VM-hypervisor isolation and gaining the hypervisor privilege. 

\mypara{Targeting victim HVM}
Compared to the paravirtual VM, a hardware-assisted VM (HVM) is unaware of underlying hypervisor and other HVMs. dFFS~\cite{razavi2016flip} demonstrates an attack against a KVM-based HVM from another HVM. It abuses the page deduplication feature to corrupt targeted files (i.e., OpenSSH public keys, Debian URLs in \texttt{sources.list} and trusted public keys in \texttt{trusted.gpg}) residing in the page cache of a victim HVM, breaking inter-HVM isolation and gaining 
full control of the victim HVM. 

\subsubsection{Leaking sensitive information}
This category of attacks aims to leak different types of sensitive information. 

\mypara{Targeting cryptographic algorithm}
{RSA, as a public-key cryptosystem, is one of the most widely used asymmetric cryptography algorithms and has been the attack target for three rowhammer attacks.}
Specifically, Bhattacharya et al.~\cite{bhattacharya2016curious} manage to induce a single bit flip fault in the secret exponent of RSA public key exponentiation in the \texttt{GNU-MP} big integer library. Similarly, JackHammer~\cite{weissman2019jackhammer} induces bit flip faults in the \texttt{WolfSSL} RSA implementation, resulting in faulty signatures. 
RAMBleed~\cite{kwong2020rambleed} is the first attack that uses rowhammer-induced bit flips as a read primitive to break inter-process isolation and leak an RSA key from the OpenSSH daemon.  

\mypara{Targeting kernel data}{
{Kernel data is a high-value target to an unprivileged attacker.} Both Spechammer (kernel exploit)~\cite{tobahspechammer} and GadgetHammer~\cite{YTobah2024gogogadget} primarily target kernel stack variables to leak kernel data. Their major difference is: Spechammer~\cite{tobahspechammer} uses rowhammer to modify attacker-controlled offset shared with the kernel for Spectre gadgets, thus discovering more Spectre gadgets and reviving Spectre attacks. For GadgetHammer~\cite{YTobah2024gogogadget}, it looks for gadgets of a particular code pattern, i.e., kernel code using nested pointer dereferences to return benign data to the user space. By flipping the pointers, arbitrary kernel data can be leaked.
}

\mypara{Targeting KASLR}
Kernel Address Space Layout Randomization (KASLR) is a Linux kernel feature that randomizes the base address of kernel's \texttt{text} segment to defend against code-reuse attacks. 
Particularly, the base address can have 512 possibilities every time the kernel boots up. While there have been many attacks that can leak KASLR (e.g., Prefetch~\cite{gruss2016prefetch}), HammerScope~\cite{cohen2022hammerscope} exploits rowhammer as a side channel to leak the run-time base address. Its key observation is that higher DRAM power consumption will increase the hammer count that is required to flip a bit. 

\mypara{Targeting DNN model}
As training a model is resource-intensive and the relevant training data is sensitive, how to protect intellectual property of a DNN model on both server and mobile platforms has attracted much attention from security researchers.
Similar to RAMBleed, DeepSteal~\cite{rakin2021deepsteal} uses the rowhammer bit flip as a read primitive and they leak weights from a target DNN model through this primitive.

\subsubsection{Degrading DNN model}
DNN has been pervasively used recently due to its stunning inference accuracy. 
Hong et al.~\cite{hong2019terminal} mount a rowhammer attack from an unprivileged process against a full-precision DNN model process. They exploit the page deduplication feature to change model weight and thus degrade model inference accuracy.
DeepHammer~\cite{yao2020deephammer}, assuming a white-box of a target quantized model, abuses the per-CPU page-frame cache update policy to corrupt specified weights in each model layer, reducing the model's  inference to random guess. 
Different from them, Li et al.~\cite{liyes} leverages the page cache eviction policy to inject faults into underlying third-party machine learning library, thus requiring no  model knowledge and achieving
model inference depletion.
\section{Rowhammer Defenses}\label{sec:defenses}

\begin{table*}[!t]
\footnotesize
\caption{A taxonomy of software defenses. For the discussion about future explorations,
please refer to \autoref{sec:insights_defenses}.}
\centering
\begin{tabular}{llll} 
\toprule
   \textbf{Objective} & 
  & \textbf{Software Defense}
  & \textbf{Requirement}
  \\ \hline
 \multirow{5}{*}{\begin{tabular}[c]{@{}c@{}} \\ \quad{Ad-Hoc} \\ \quad{Attempt}\end{tabular}}  
  & \multirow{2}{*}{double DRAM row refresh rate}      & Computer manufacturers
   & \multirow{2}{*}{BIOS updates}   
   \\ 
  &     & ~\cite{Apple,HP,LENOVO}    
  &     \\\cline{2-4}
  & {disallow cache-flush instructions }      & Chrome browser  &  \multirow{2}{*}{Browser sandbox updates}
  \\ 
  & {disallow non-temporal instructions}     & ~\cite{nonclflush,nontemporal}  &
  \\ \cline{2-4} 
  & remove non-root access to \texttt{pagemap} & {Linux kernel~\cite{shutemovpagemap}}  & \multirow{1}{*}{Kernel updates}
  \\ 
  & disable page deduplication &  {Clouds~\cite{pagededupdisabled}} & Hypervisor updates
  \\ \midrule
\multirow{6}{*}{\begin{tabular}[c]{@{}c@{}}\\ \\{DRAM-Aware}\\ {Reinforce}\end{tabular}}  
  & user-kernel isolation        & CATT~\cite{brasser17can} &  \multirow{5}{*}{Kernel updates} 
  \\ 
  & inter-process isolation      & {RIP-RH~\cite{bock2019rip}}  &
  \\ 
  & \multirow{2}{*}{DMA buffer isolation}   & \multirow{1}{*}{ALIS~\cite{tatar2018throwhammer}}  &
  \\
    &  &\multirow{1}{*}{GuardION~\cite{van2018guardion}} &
    \\ 
   & page-table isolation         & {CTA~\cite{wu2018CAT}}  &
   \\ \cline{2-4}
  & intra-HVM isolation         & {ZebRAM~\cite{konoth2018zebram}}  &    Kernel \& hypervisor updates
  \\ \cline{2-4}
   & \multicolumn{1}{l}{\textbf{inter-HVM isolation}} &  \multicolumn{2}{l}{\multirow{2}{*}{\textbf{Future Exploration}}}  \\
   & \multicolumn{1}{l}{\textbf{HVM-hypervisor isolation}} &    \\ \midrule
\multirow{7}{*}{\begin{tabular}[c]{@{}c@{}}{RH-Triggered}\\{~Detect}\end{tabular}} 
  &  abnormal CPU cache misses   & \multirow{2}{*}{ANVIL~\cite{aweke2016anvil}} & \multirow{2}{*}{Add-on kernel module}
  \\
  &  abnormal memory-access patterns   &  & 
  \\ \cline{2-4}
  &  abnormal EM emanations  & {RADAR~\cite{zhangleveraging}}   &  Software-defined radio device
  \\ \cline{2-4}
  &  abnormal binary code & {MASCAT~\cite{irazoqui2016mascat}}  &   User-space static analysis tool
  \\ \cline{2-4}
  &  abnormal row-access patterns & {SoftTRR~\cite{zhang2021softtrr}}  &   Add-on kernel module
  \\ \cline{2-4}
  &  \textbf{abnormal DRAM power} &  \multicolumn{2}{l}{\multirow{2}{*}{\textbf{Future Exploration}}}  \\ 
  &  \textbf{abnormal row-buffer conflicts} &  \\ 
\bottomrule
\end{tabular}
\label{tab:software_defenses}
\end{table*}

\subsection{Software Defenses}
Existing software defenses are proposed to defend commodity systems against rowhammer attacks and they can be classified into three categories, i.e., {ad-hoc attempts}, {DRAM-aware isolation} and RowHammer(RH)-triggered detection, shown in \autoref{tab:software_defenses}. Each of them aims to mitigate step \circled{1} or \circled{2} in the methodology component of~\autoref{fig:attack_framework}. In the following, we discuss each category in detail. 

\subsubsection{Ad-Hoc Attempts}\label{sec:ad_hoc}
We introduce four major countermeasures in this category. 
\emph{First}, multiple computer manufacturers such as Apple~\cite{Apple}, HP~\cite{HP}, and Lenovo~\cite{LENOVO} provide BIOS updates to double the DRAM refresh rate, that is, the DRAM refresh period is decreased from 64\,ms to 32\,ms. However, ANVIL~\cite{aweke2016anvil} still can induce bit flips in the doubling refresh rate. Even worse, Kim et al.~\cite{kim2020revisiting} observe that a bit flip can occur in mainstream DDR4 chips within 1\,ms. 
\emph{Second}, Google chrome browser has disallowed the use of both explicit cache-flush instructions~\cite{nonclflush} and non-temporal instructions~\cite{nontemporal}, mitigating the NaCl exploit from~\cite{seaborn2015exploiting}. Still, there are other available techniques in avoiding CPU cache in ~\autoref{tab:techniques}, e.g., cache eviction and direct memory access.
\emph{Third}, since version 4.0, Linux kernel removes non-root access to the \texttt{pagemap} interface~\cite{shutemovpagemap}, which prevents potential attackers from acquiring the mapping from a virtual address to a physical address. Clearly, the countermeasure is not effective, as discussed in \autoref{sec:bypass_rowbuffer}. 
\emph{Last}, the {page deduplication} feature has been turned off by default only to mitigate specific attacks~\cite{razavi2016flip, bosman2016dedup} in commodity cloud platforms~\cite{pagededupdisabled}.

\subsubsection{DRAM-Aware Domain Isolation}
This category of defenses separates security-sensitive memory from attacker-accessible memory in DRAM by putting guarding rows in between. By doing so, the attacker-induced bit flips are absorbed by the guarding rows and will not affect target security domain. {Existing schemes reinforce MMU-enforced isolation at different granularities to protect different security domains~\cite{brasser17can,bock2019rip,konoth2018zebram,tatar2018throwhammer,van2018guardion,wu2018CAT}.}

\mypara{Reinforcing user-kernel isolation}
To defend against rowhammer attacks targeting the kernel, CATT~\cite{brasser17can} partitions the DRAM rows within each bank into two parts-- one part for the user domain and the other part for the kernel domain, and uses at least one empty row to separate the two domains. As the empty rows will absorb the bit flips from the user domain,
the kernel's integrity is protected and little runtime overhead (within 1\%) is incurred.

\mypara{Reinforcing inter-process isolation}
To prevent rowhammer attacks targeting security-sensitive process such as \texttt{setuid}, RIP-RH~\cite{bock2019rip} enforces DRAM-aware memory isolation for the targeted processes by segregating their physical memory into dedicated DRAM areas.
As such, these processes are not allowed to share memory with other processes, otherwise, the attacker can hammer the shared memory. Their evaluated benchmarks report a modest run-time overhead (within 4\%).

\mypara{Reinforcing DMA buffer isolation}
There are two reinforced DMA isolation schemes, that is, 
ALIS~\cite{tatar2018throwhammer} on x86 and GuardION~\cite{van2018guardion} on ARM.
ALIS surgically isolates RDMA-enabled DMA memory with guarding rows and thus every attacker-generated bit flip is confined to guarding rows, defending against Throwhammer~\cite{tatar2018throwhammer} and introducing almost no run-time overhead to the protected system. Similar to ALIS, GuardION~\cite{van2018guardion} mitigates DMA-based attacks particularly on mobile devices by isolating ION-enabled DMA buffers using guarding rows. Also, its overall performance overhead is acceptable, i.e., within 6\% on average performance increase and 7\% decrease in its worst case.

\mypara{Reinforcing page-table isolation}
As all existing kernel-privilege-escalation attacks target corrupting page tables, CTA~\cite{wu2018CAT} aims to reinforce DRAM-aware page table isolation. It proposes a two-step approach. In the first step, 
page-table pages are allocated from a dedicated memory region that has high physical addresses and resides at the end of the memory space. 
In the second step, rows with true cells are selected from the memory region for page-table allocation. As the dedicated memory region is not separated from the rest of memory region with guarding rows, a bit flip can occur in the page-frame-number field of a page-table entry (PTE). However, the corrupted PTE will only point to a new physical address lower than the original one and it will never point to another PTE. Thus, the attacker cannot gain access to a page-table page. Base on their evaluated benchmarks, little overhead (within 1\%) is reported. 

\mypara{Reinforcing Intra-HVM Isolation}
ZebRAM~\cite{konoth2018zebram} leverages hardware assisted hypervisor to {split} DRAM rows of a target HVM into two regions in a zebra pattern, that is, even rows for safe system data and odd rows for unsafe swap space. Thus, hammering even rows is effective in inducing bit flips in adjacent odd rows of swap space. However, hammering odd rows is too slow to incur a single bit flip in adjacent even rows where system data reside.
For potentially corrupted data in the swap space, ZebRAM performs data integrity check before using them. Its evaluated benchmark reports an overhead of 5\%.

\subsubsection{RH-Triggered Detection}\label{sec:software_detection}
For step \circled{2} in methodology of~\autoref{fig:attack_framework}, existing rowhammer attacks require hammering target objects, which results in abnormal software/hardware side effects. We note that MAD~\cite{wiesingermad} can detect many rowhammer attacks that have abnormal behaviors in memory allocations. However, it has not been evaluated on a commodity system (e.g., security effectiveness and performance overhead) and thus is not listed in \autoref{tab:software_defenses}.

\mypara{Detecting abnormal CPU cache misses and memory-access patterns} 
As a large majority of attacks have two characteristics (i.e., high CPU cache misses and high spatial locality of memory accesses),  
ANVIL~\cite{aweke2016anvil} proposes a two-step approach based on Intel performance monitoring counters. First, it monitors CPU last-level cache (LLC) miss rate. If the miss rate becomes greater than a predefined threshold, it then starts to sample physical addresses of memory accesses that miss the LLC. If the sampled addresses exhibit a high temporal locality, i.e., in the same DRAM bank, it will refresh rows adjacent to rows of the sampled addresses. 
Its overall runtime overhead is low within 1\% but it incurs a worse-case overhead of 8\%. 

\mypara{Detecting abnormal binary code}
Most rowhammer attacks use explicit cache-flush instructions, non-temporal instructions or cache eviction sets to avoide CPU caches. With this observation, Irazoqui et al.~\cite{irazoqui2016mascat} implement a static analysis tool, coined MASCAT, to identify the typical rowhammer attributes within target binary code. They note that the tool is an extensible framework and can include other rowhammer attributes. As this tool performs a static analysis, it does not introduce run-time overhead.

\mypara{Detecting abnormal EM emanations}
With a key observation that hammering activities emanate distinguishable electromagnetic (EM) signals, RADAR~\cite{zhangleveraging} leverages a \$299 radio-based external device to capture EM signals from DRAM. After the EM emanations are processed, RADAR can expose recognizable hammering-correlated sideband patterns in the spectrum of the DRAM clock signal. To achieve effectiveness and robustness classification, RADAR feeds hammering-correlated spectrograms into a convolutional neural network (CNN) for training. If the CNN's inference indicates an ongoing rowhammer activity, RADAR can kill suspicious processes. Since it works outside of the protected system, it imposes almost no overhead to the system.

\mypara{Detecting abnormal row-access patterns}
SoftTRR~\cite{zhang2021softtrr} is a software target-row-refresh technique that prevents page tables from being corrupted. It leverages the \texttt{rsrv} bits in page table entries to frequently track accesses to any rows that are adjacent to rows hosting page tables.
When the tracked access counter reaches a pre-determined threshold, refreshes will be performed to corresponding rows with page tables.
Its runtime overhead is negligible within 1\%.

\subsubsection{Summary of Software Defenses}
Clearly, the ad-hoc attempts have security issues that are discussed in \autoref{sec:ad_hoc}. Below, we discuss the major limitations of the other two categories in \autoref{tab:software_defenses}.

\mypara{DRAM-aware domain isolation schemes}
\emph{First,} all of the schemes defend against multiple (not all) {row-hammer} rowhammer attacks. Besides, they require modifying the memory allocator either in the kernel or the hypervisor to enforce DRAM-aware memory allocation, making them hard to adopt in commodity systems. \emph{Second,} {as rowhammer affects non-adjacent rows, the blast radius from a hammered row is dependent on specific DRAM chips.} From the observation~\cite{kim2020revisiting}, the blast radius of some chips can be up to 6. Thus, the defenses need to determine the blast radius for given chips and waste the rows within the  blast radius as the guarding rows or the swap-space memory. When the blast radius grows, the wasted rows increases. 
\emph{Last,} all these defenses except CTA~\cite{wu2018CAT} rely on the invalid assumption that logical and physical DRAM addresses are the same, ignoring the DRAM row remapping. For CTA, its security guarantee is broken by the data-scrambling feature enforced by the memory controller~\cite{he2023whistleblower}.

{\mypara{RH-triggered detection schemes}}
RADAR~\cite{zhangleveraging} requires a specialized radio device, which is unlikely to be purchased and deployed in practice by individual laptops, company servers, and clouds. \emph{Besides,} both ANVIL~\cite{aweke2016anvil} and MASCAT~\cite{irazoqui2016mascat} cannot detect attacks that originate from non-CPU hardware components, e.g., GLitch~\cite{frigo2018grand}. SoftTRR~\cite{zhang2021softtrr} implementation is ineffective in cases where the rowhammer blast radius is greater than 6. 

\begin{table}[ht]
\footnotesize
\caption{A taxonomy of {production} defenses deployed in commodity DRAM. } 
\centering
\resizebox{1.0\columnwidth}{!}{
\begin{tabular}{lll} 
\toprule
\multirow{2}{*}{\textbf{Objective}} & \multirow{2}{*}{{\textbf{Production Defense}}} & \multirow{2}{*}{\textbf{Prototype Location}}  \\ 
            & & \\ \hline
\multirow{4}{*}{\begin{tabular}[c]{@{}c@{}}{Data Integrity Check} \\{(Correct/Detect bit flips)}\end{tabular}} 
                & \multirow{2}{*}{\begin{tabular}[l]{@{}l@{}}{MC-Aware DRAM-based ECC~\cite{ryan2009channel}} \\ \end{tabular}}       &  \multirow{2}{*}{MC \& Extra DRAM}\\ 
                                                            &                                 &              \\ \cline{2-3}
                                                            & On-die ECC~\cite{nair2016xed}               & \multirow{2}{*}{DRAM}     \\ 
                                                            & \textbf{(Future Exploration*)} &    \\ 
               
\midrule
\multirow{2}{*}{\begin{tabular}[c]{@{}c@{}}{Row Activation Count}\\{(Refresh specific rows)}\end{tabular}}           
               & \multirow{2}{*}{on-die TRR~\cite{DDR4TRR}}   & \multirow{2}{*}{DRAM}  \\
               &                                              & \\
              
\bottomrule
\end{tabular}
\label{tab:production_defenses}
}
\end{table}

\subsection{{Production} Defenses}\label{sec:product}
Here, we classify existing {production} defenses into two categories, shown in~\autoref{tab:production_defenses}, which are introduced below (Section Appendix provides a more detailed discussion).

\subsubsection{Data {Integrity Check}}\label{sec:integrity_check}
This category uses different algorithms to detect and correct bit flips by performing data integrity check. 

\mypara{MC-Aware DRAM-based ECC}
ECC (Error Checking and Correcting) is proposed to mitigate data-corruption by introducing additional bits to store the parity of data that is written into DRAM. A {production-level} case of ECC is single-error-correction-double-error-detection (SECDED) hamming code. In this case, ECC functions are implemented in the {MC} on commodity processors, that is, the MC has 8-bit hamming SECDED code with every 64-bit data. Thus, the data bus between DRAM and MC is extended from 64 bits to 72 bits and each {rank} has an extra chip to store ECC code. 

\mypara{On-die ECC}
Different from the aforementioned ECC, on-die ECC, as its name suggests, is integrated directly into the same DRAM chips and invisible to the MC~\cite{nair2016xed,patel2019understanding}. 
As the on-die ECC operates entirely within the same DRAM chips, it is adopted by mainstream DRAM manufacturers to reduce the data-corruption rates in new generations of DRAM chips (e.g., LPDDR4~\cite{oh20143} and {DDR5}~\cite{ondieeccddr5}). As the DRAM manufacturers consider the on-die ECC functions as proprietary, its details cannot be found in public documentation and have been reverse-engineered in recent works~\cite{patel2019understanding,patel2020bit,patel2021harp}.

\subsubsection{Row Activation Count}\label{sec:row_act}
DRAM manufacturers implement different schemes of row-level on-die Target Row Refresh (TRR)~\cite{DDR4TRR} to prevent data in DRAM from being flipped. {As its name suggests,} TRR counts rows that are being activated and refreshes their adjacent rows when the counts reach a predefined threshold. Typically, a TRR scheme implements a bank-level counter table, and counts a limited number (e.g., 16) of different rows within the same bank.

\subsubsection{Summary of {Production} Defenses}
All existing {production} defenses except the on-die ECC have been proved vulnerable to rowhammer. Specifically, the MC-Aware DRAM ECC was believed to deter rowhammer-based attacks until ECCploit~\cite{cojocar2019exploiting} in 2019 demonstrated that it was susceptible to rowhammer by inducing enough (e.g., 3) bit flips in a single code word (e.g., 72 bits). 

{For on-die ECC, given its functions have been reverse engineered as discussed above, \emph{it is worth investigating whether it is also susceptible to rowhammer.}
For on-die TRR implementations in recent DRAM, they are publicly undocumented. However, different TRR implementations from three major DRAM manufacturers have been reverse-engineered recently and DRAM based on them are still vulnerable to rowhammer~\cite{frigo2020trrespass,hassan2021uncovering,kogler2022half}, breaking their security guarantee. 
\section{Our Insights on Future Explorations}\label{sec:insights}
In this section, we first discuss attacks followed by a DRAM-aware memory isolation in commercial clouds. We then present how to detect rowhammer motivated by its key characteristics.

\subsection{Potential Rowhammer Attacks}\label{sec:insights_attacks}
\mypara{Gaining hypervisor privilege}
The only rowhammer attack against HaV employed by public clouds is from Razavi et al.~\cite{razavi2016flip}. However, this attack has been rendered unavailable since page deduplication  was disabled. Also, the attack achieves a control of a
target HVM without compromising the hypervisor.
Thus, it is worth exploring rowhammer to break MMU-enforced HVM-hypervisor isolation and gain hypervisor privilege in public clouds. 

In native environments, MMU maintains a single layer of page tables, which maintains the mappings between virtual addresses and physical addresses. 
In HaV, mainstream microarchitecture vendors (e.g., Intel~\cite{intelvt} and AMD~\cite{amdvt}) modifies MMU to maintain two layers of page tables. The first-layer page table, i.e., Guest Page Table (GPT), is managed by HVM, and the second layer, i.e., Extended/Nested Page Table (EPT/NPT), is managed by the hypervisor. 
For a memory access from the HVM, MMU performs page-table walk at both levels (if combined TLB and host TLB searches miss) and retrieve relevant page-table entries (PTEs) from memory (if PTEs are cached). If the attacker can produce frequent implicit accesses to memory locations where EPTs/NPTs reside, they can corrupt other adjacent EPTs and gain the hypervisor privilege. 

\mypara{Exploiting isolated GPUs in x86}
As isolated GPUs in the x86-based processors are widely used for training DNN model tasks, mainstream cloud providers (e.g., Amazon, Google, Oracle, and Alibaba) offer cloud GPU services in a multi-tenant manner~\cite{cloudgpus}. With this service, individuals or small businesses are charged for the time they have used the shared cloud GPUs for their DNN model training instead of purchasing their own physical GPUs. 
Thus, in a real-world scenario where an attacker shares cloud GPUs and GPU memory with a victim tenant, it is worth exploring whether the attacker can refrain from using GPU caches, induce bit flips in GPU memory of a special DRAM type (e.g., GDDR5 is vulnerable to rowhammer~\cite{LENOVO}) and mount end-to-end attacks, such as compromising the DNN model of a victim tenant. 

\subsection{Potential Rowhammer Defenses}\label{sec:insights_defenses}
\mypara{Reinforcing DRAM-aware isolation in hardware-assisted virtualization} 
To counteract the aforementioned cloud-based rowhammer candidates, we can reinforce a DRAM-aware isolation for different security domains, such as inter-HVM and HVM-hypervisor. As kernel-based virtual machine (KVM) in the Linux kernel is widely used as a hypervisor, the isolation prototype can be based on KVM by extending its physical memory allocator to be rowhammer-aware. This KVM-based isolation is expected to be transparent to the HVMs and incur small performance overhead. A good example is Siloz~\cite{loughlin2023siloz} that was published recently. It enforces DRAM-aware isolation in KVM by placing each VM’s and hypervisor’s data onto different DRAM subarray groups with guarding rows in between. 

\mypara{Detecting abnormal DRAM power} 
To improve performance, the processor always tries to serve memory-access requests with its caches. For rowhammer attacks, they induce frequent memory accesses to DRAM, which is expected to exhibit an abnormal distribution of power consumption from DRAM over a specific time period, thus being a possible solution to rowhammer detection.
Intel running average power limit (RAPL) may be leveraged. Specifically, it is a mechanism of setting power and thermal limits on the processor packages and DRAM~\cite{intelOp}, which is available since Sandy Bridge. 
{Intel RAPL has a set of model-specific registers (MSRs) to monitor the power consumption over a short time interval for different domains, e.g., package, power planes, and DRAM.
For user-space access, Intel implements a power capping framework (i.e., \texttt{powercap}), which exposes the MSRs through \texttt{sysfs}.} Note that since the RAPL-based power side-channel has been discovered~\cite{lipp2021platypus}, only root-privilege users have legitimate access to the \texttt{sysfs} interface.
As such, it is worth investigating Intel RAPL to detect abnormal DRAM power induced by rowhammer.

\mypara{Detecting abnormal row-buffer conflicts} 
Given that rowhammer requires accessing one or multiple rows frequently to trigger bit flips, it will result in an abnormal number of row buffer conflicts, which is thus an indicator of an ongoing rowhammer activity. 
We observe that an Intel server can be of help by providing the statistics of row buffer conflicts based on the Intel manual~\cite{xeon}.
Specifically, there exists a critical part outside Intel cores and it is called ``Uncore". The uncore part includes PCI-express, memory controller, etc. Intel implements a list of performance counter events to monitor the performance of the uncore part, among which the \texttt{PRE\_COUNT.PAGE\_MISS} event counts DRAM \texttt{PRE} events caused by \texttt{page misses}~\cite{xeon}. Here, a page miss occurs when a row buffer is open but has a wrong row in it and thus it is resulted from a ``row-buffer conflict". 
Besides \texttt{PRE\_COUNT.PAGE\_MISS}, the \texttt{CAS\_COUNT.RD} event counts all DRAM read requests and the \texttt{CAT\_COUNT.WR} event records all DRAM write requests. To this end, 
another event called \texttt{PCT\_REQUESTS\_PAGE\_MISS} reports the percentage of memory requests that result in row buffer conflicts, i.e., \texttt{PRE\_COUNT.PAGE\_MISS / (CAS\_COUNT.RD + CAS\_COUNT.WR)}.
Consequently, {how to leverage} {\texttt{PCT\_REQUESTS\_PAGE\_MISS}} to detect rowhammer activities in real-time is worth investigating. 
\section{Conclusion}\label{sec:conclusion}
Motivated by Kim et al.~\cite{kim2014flipping} in 2014, 
a large number of rowhammer attacks have been proposed against applications, kernels and clouds. As a response, software and hardware defenses from both academia and industry are developed for commodity systems. As rowhammer is unlikely to be addressed in the near future, this work can help the community deeply understand rowhammer on commodity systems with production DRAM chips deployed. We also believe new attack vectors and countermeasures (besides \autoref{sec:insights}) will be developed to study the causes and impacts of rowhammer in real-world.

\begin{acks}
We would like to thank the anonymous reviewers for their valuable feedback. This research was supported partly by the SCUT Research Startup Fund No. K3200890. Z.~Zhang and D.~Chen contribute equally. Y.~Gao is the corresponding author.
\end{acks}

\bibliographystyle{ACM-Reference-Format}
\bibliography{defs,main}

\appendix
\label{sec:appendix}
\begin{table*}[ht]
\footnotesize
\caption{A taxonomy of hardware defenses. (Whether on-die ECC is effective against rowhammer has not been studied.)} 
\label{tab:hardware_defenses}
\centering
\begin{tabular}{llll} 
\toprule
\multicolumn{2}{c}{\textbf{Objective}} & {\textbf{Hardware Defense}}  & \textbf{Prototype Location}   \\ \hline
\multirow{6}{*}{\begin{tabular}[c]{@{}c@{}}{Data Integrity} \\{Check}\end{tabular}} 
                & \multirow{3}{*}{Correct/Detect bit flips}   & {MC-Aware ECC~\cite{ryan2009channel}, SafeGuard~\cite{fakhr2022safeguard}}       &  DRAM and MC\\ \cline{3-4}
                &&{CSI~\cite{juffinger2022csi}}, {PT-Guard~\cite{saxena2023ptguard}}& {MC} \\ 
                \cline{3-4}
                &                                             & On-die ECC~\cite{nair2016xed} \textbf{(Future Exploration*)}               & {DRAM}     \\ \cline{2-4}
                & \multirow{2}{*}{Detect bit flips}           & Dynamic Skewed Hash Tree~\cite{vig2018rapid}   & MC                      \\\cline{3-4}
                &                                             & Dummy Cells~\cite{gomez2016dram}          & DRAM    \\ 
\midrule
\multirow{12}{*}{\begin{tabular}[c]{@{}c@{}}{Row Activation}\\{Count}\end{tabular}}
               & \multirow{9}{*}{Refresh specific rows}    
               & on-die TRR~\cite{DDR4TRR}, Panopticon~\cite{bennett2021panopticon}, TWiCe~\cite{lee2019twice}
               & \multirow{3}{*}{DRAM}           \\  & &Bains et al.~\cite{bains2014distributed}, Fisch et al.~\cite{fisch2017dram}, Kim~\cite{thereof} & 
               \\  & &Greenfield et al.~\cite{greenfield2014method}, ProTRR~\cite{marazzi2022protrr} &
               \\\cline{3-4}
               &                                           & CAT~\cite{seyedzadeh2018mitigating}, Graphene~\cite{park2020graphene}  , Bains et al.~\cite{bains2015method,RRC}, & \multirow{2}{*}{MC}              \\ & &Cowles et al.~\cite{same}, Greenfield et al.~\cite{greenfield2015row}& \\
               \cline{3-4}
               &                                           & CRA~\cite{kim2014architectural}                          & DRAM and MC          \\ \cline{3-4}
               &                                           & Mithril~\cite{kim2021mithril}                         & DRAM and/or MC  \\
               \cline{3-4}
               &                                           & Silver Bullet~\cite{devaux2021method,yauglikcci2021security}, Bains et al.~\cite{bains13row}             & {DRAM or MC}\\
                \cline{2-4}
               & {Throttle specific row activations}    & BlockHammer~\cite{yauglikcci2021blockhammer}, Greenfield et al.~\cite{greenfield2016throttling}     & {MC} \\  \cline{2-4}
               
             & \multirow{2}{*}{{{Swap aggressor rows with random rows}}}
             & Saileshwar et al.~\cite{saileshwar2022randomized},
             & \multirow{2}{*}{{DRAM and MC}}  \\
             && Woo et al.~\cite{woo2023ssrs}, SHADOW~\cite{wi2023shadow}&  \\ 
\midrule
                                                          
\multirow{2}{*}{\begin{tabular}[c]{@{}c@{}}{Row-Activation-Triggered}\\{Probabilistic Refresh}\end{tabular}}         
              & \multirow{2}{*}{Refresh specific rows probabilistically}                   & \multirow{2}{*}{PARA~\cite{kim2014flipping}, PRoHIT~\cite{son2017making},
              MRLoc~\cite{you2019mrloc}}& \multirow{2}{*}{MC} 
              \\  \\
\hline
\multirow{2}{*}{\begin{tabular}[c]{@{}c@{}}{Cache Miss/Flush and}\\{DMA Access Count}\end{tabular}}         
              & \multirow{1}{*}{Relocate targeted data and/or}                   & \multirow{2}{*}{LightRoAD~\cite{taouil2021lightroad}}             & \multirow{2}{*}{MC}           \\ 
              &  \multirow{1}{*}{Disable DMA access temporarily}    &   & \\
\hline
\multirow{2}{*}{\begin{tabular}[c]{@{}c@{}}{Fabrication Process}\\{Improvement}\end{tabular}} 
             & \multirow{1}{*}{{Mitigate electromagnetic coupling }}                  & Yang et al.~\cite{yang2016suppression}, Gautam et al.~\cite{gautam2018improvement, gautam2019row, 9025775}                        & {DRAM}   \\\cline{2-4}
             &    Reduce acceptor-type traps                                          &  Ryu et al.~\cite{ryu2017overcoming}                         &   DRAM                               \\
\bottomrule
\end{tabular}
\end{table*}

\section{Hardware Defenses of Rowhammer}
{In this section, we analyze existing hardware-based defenses of rowhammer from both academia and production.}
As shown in ~\autoref{tab:hardware_defenses}, we classify existing hardware defenses into multiple categories, which are introduced below.

\subsection{Data {Integrity Check}}
This category of defenses uses different algorithms to detect or even correct bit flips by performing data integrity check. 

\mypara{MC-Aware DRAM-based ECC}
As cosmic rays or alpha particles~\cite{hwang2012cosmic,lantz1996soft} can corrupt data in DRAM chips accidentally, ECC is proposed to address the data-corruption problem by storing additional parity bits in separate chips next to the data bits in the original chips. {When data is accessed, ECC calculates its parity and compares it against the stored one.}
Theoretically, ECC can correct $n$ ($n \geq 1$) bit flips and detect more than $n$ bit flips. 

An production-level use case of ECC is single-error-correction-double-error-detection (SECDED) hamming code. In this case, ECC functions are implemented in the {MC} on commodity processors, that is, the MC has 8-bit hamming SECDED code with every 64-bit data. Thus, the data bus between DRAM and MC is extended from 64 bits to 72 bits and each {rank} has an extra chip to store ECC code.
This MC-Aware DRAM ECC was believed to deter rowhammer-based attacks until ECCploit~\cite{cojocar2019exploiting} in 2019 demonstrated that it was susceptible to rowhammer by inducing enough (e.g., 3) bit flips in a single code word (e.g., 72 bits). 

To reinforce the MC-Aware DRAM-based ECC against rowhammer, a couple of new defenses~\cite{kim2019effective,wang2019reinforce} have been proposed. Particularly, they use row-remapping schemes to distribute bit-flips to different rows, which significantly reduces the occurrences where undetectable multiple bit-flip errors occur in a single word.
{Besides, cryptographic Lightweight Message Authentication Codes (MAC) are stored in the ECC memory or MC to support detect and correct rowhammer-induced bit flips~\cite{fakhr2022safeguard,juffinger2022csi,saxena2023ptguard}. }
Further, the design principle of {future} ECC is expected to put more weight on its detection capability rather than its correction capability~\cite{qureshi2021rethinking}, thus achieving a better detection of rowhammer.

\mypara{On-die ECC}
As introduced in~\autoref{sec:product}, on-die ECC, is implemented in the DRAM chips and invisible to the MC~\cite{nair2016xed,patel2019understanding}. 
It is adopted by mainstream DRAM manufacturers to reduce the data-corruption rates in recent DRAM chips (e.g., LPDDR4~\cite{oh20143} and {DDR5}~\cite{ondieeccddr5}). While the on-die ECC functions are publicly unavailable, they have been reverse-engineered~\cite{patel2019understanding,patel2020bit,patel2021harp}.

\mypara{Dynamic Skewed Hash Tree}
Vig et al.~\cite{vig2018rapid} propose a lightweight scheme within the MC to check data integrity. Particularly, they apply a sliding window protocol to identify potentially corrupted victim rows {and use} a dynamic integrity tree structure with SHA-3 Keccak hash functions for bit-flip detection.

\mypara{Dummy Cells}
Gomez et al.~\cite{gomez2016dram} propose the so-called \emph{dummy cells} to perform data integrity checks. Compared to regular cells, the dummy cells are intrinsically more susceptible to rowhammer as they have smaller capacitance and larger transistors. Every row will contain a dummy cell that is fully charged during the DRAM refresh time. If the MC reads out a row with its dummy cell value bit-flipped, it indicates an early warning of data loss and thus the MC can refresh it and its adjacent rows. 

\subsection{Row Activation Count}
In this category, defenses rely on hardware counters to count the number of activations per row within a DRAM refresh period. If the count for a row exceeds a predefined threshold, they will refresh its adjacent rows, or throttle further activations to the row, or swap the row with another random row. We discuss the defenses based on where the counters are implemented in the DRAM hierarchy.

\mypara{Row-Level Count with One Counter per Row}
CRA~\cite{kim2014architectural}, as an early scheme published in 2014, maintains a counter per row using a portion of the DRAM memory. In such a case, every time a row is activated, its counter needs to be updated, resulting in doubled memory-access latency. To reduce the latency, CRA implements a cache of recently updated counters within the MC. Thus, a row counter is accessed from DRAM {if} it cannot be found from the cache. Similar to CRA, Panopticon~\cite{bennett2021panopticon}, a recent scheme in 2021, also maintains a counter in DRAM for each row. Unlike CRA, Panopticon rearranges DRAM's subarrays' layout in a staggered way, and thus {a counter can be updated alongside a row activation}.

\mypara{Row-Level Count with One Counting Logic per Bank}
Besides the on-die Target Row Refresh (TRR)~\cite{DDR4TRR} introduced in~\autoref{sec:product}, the following works from the academia are also in this category.

TWiCe~\cite{lee2019twice} manages a counter table for each DRAM bank in the register clock driver (RCD), which is inside a DIMM but separated from chips. Each entry of the counter table comprises of a valid bit, a row address, an activation counter, and a lifetime counter. When a row is being activated, TWiCe increments its activation counter if it exists in the table, otherwise allocates a spare entry. If the counter reaches a predefined threshold, its adjacent rows are refreshed. For every periodic DRAM refresh command, TWiCe starts to prunes the table. When a row's average activation rate is lower than a specific threshold during its lifetime, its entry is pruned with its valid bit set to 0. After each pruning, the remaining entries' lifetime counters are incremented. Whenever a row's activation count reaches the threshold value, TWiCe will issue a new command called Adjacent Row Refresh (ARR) to refresh the adjacent victim rows. Whenever a row's activation count reaches the threshold value, TWiCe will issue a new command called Adjacent Row Refresh (ARR) by extending the DRAM protocol to inform the DRAM of the aggressor row's address, which will then refresh the adjacent victim rows. 

Graphene~\cite{park2020graphene}, implemented in the MC, maintains a counter table and a spillover counter for each DRAM bank. Each entry of the counter table uses a row address as the key and an estimated counter as the value. The spillover counter is a special register and its value represents the upper bound of \dramcommand{ACT} command counts for all rows which are currently not in the table. Graphene uses the Misra-Gries algorithm to track aggressor rows. Specifically, upon a row activation, if the row address is already in the table, the relevant estimated counter is incremented by one. If the address is not in the table and the table is not full, it will be inserted into the table with the estimated counter set to one. If the table is full, Graphene first checks whether there is an entry whose estimate counter is equal to the spillover counter. If there is, it replaces this entry with the row address that needs to be inserted and increments the existing estimated counter by one. Otherwise, it skips the row address by simply increasing the spillover counter by one. In order not to miss more new row addresses, both the counter table and spillover counter are reset periodically. Within each reset window, if a row's estimated counter reaches a threshold, Graphene will issue a new nearby row refresh (NRR) command by extending the DRAM protocol to refresh the row's adjacent rows. To prevent non-adjacent rowhammer bit flips, Graphene can modify the number of rows that the NRR command covers. 

Similarly, Mithril~\cite{kim2021mithril} leverages a newly introduced DRAM command called refresh management (RFM) in DDR5~\cite{ondieeccddr5} to implement a customized Misra-Gries algorithm. ProTRR~\cite{marazzi2022protrr} proposes an in-DRAM Misra-Gries algorithm to optimize TRR. Unlike Graphene, these two defenses do not require any change to the DRAM protocol. 

Instead of refreshing the potential victim rows in advance, BlockHammer~\cite{yauglikcci2021blockhammer} proactively throttles memory accesses that are considered malicious. Implemented in the MC, BlockHammer contains two components: RowBlocker and AttackThrottler. RowBlocker employs two counting bloom filters per bank to alternately track the activation rates of all DRAM rows in a rolling time window. If a row’s activation rate exceeds a predefined threshold, RowBlocker will blacklist this row. RowBlocker also maintains a first-in-first-out (FIFO) history buffer per rank to record rows that have been activated in the last fixed time window of $t_{Delay}$. If a row is not only blacklisted but also in the history buffer (i.e., recently activated), RowBlocker will block any further activations to this row for $t_{Delay}$ to prevent the row from being further hammered. Based on the results from RowBlocker, AttackThrottler cuts the memory bandwidth provided to threads that are identified as potential attackers, thus allowing co-running benign threads to have higher performance when accessing memory. 

Similarly, most existing patents count row activations in a specified time period (e.g., \texttt{tREFI}) by maintaining a counting logic per memory portion (e.g., DRAM bank). If a row's activation count exceeds a predefined threshold, either its victim rows will be refreshed~\cite{greenfield2014method,bains2015method,bains2014distributed,bains13row,greenfield2015row,fisch2017dram,RRC}, or subsequent \dramcommand{ACT} commands sent to the row will be throttled~\cite{greenfield2016throttling}. 

Besides refresh- and throttling- based defenses, Saileshwar et al.~\cite{saileshwar2022randomized} propose to swap aggressor rows and other random rows when row activation count reaches a predefined threshold. By doing so, they can break the physical proximity between the aggressor row and its original victim rows, thus mitigating rowhammer, so-called \emph{row shuffle}. Further, Woo et al.~\cite{woo2023ssrs} and SHADOW~\cite{wi2023shadow} respectively enhance the security and performance of the row-shuffle defense.

\mypara{RowSet-Level Count}
CAT~\cite{seyedzadeh2018mitigating} applies one counter to a set of rows and every access to rows within the set will be counted. 
However, in an intuitive implementation where rows are evenly assigned to each counter, some counters can become quite ``busy" and some remain ``idle", as benign workloads in production computing environments have memory access locality and specific memory regions of rows are accessed more often (``hot") than other memory regions. To address this problem, CAT leverages \emph{adaptive trees} to dynamically update the number of rows one counter manages. By doing so, more counters are allocated for a ``hot" memory region and less counters are for a ``cold" memory region. 

The following patents also use a similar strategy of the rowset-level count. Specifically, Silver Bullet~\cite{devaux2021method,yauglikcci2021security} assigns a counter to a subbank (i.e., a rowset), size of which can be tuned from two times the blast radius to the row number of a bank. We note that a subbank size by Silver Bullet is decided when designing a DRAM chip and cannot change thereafter, which is different from CAT. Every time a counter reaches a prefixed value, Silver Bullet will refresh one row in the corresponding subbank and clear the counter value to zero. Silver bullet refreshes rows in a subbank one by one in a round-robin way. Kim~\cite{thereof} assigns a counter to each DRAM mat (e.g., a subarray), and rows in a mat are refreshed when its counter reaches a predetermined value. Cowles et al.~\cite{same} assign a counter to a bank (a rowset here thus refers to all the rows in a bank) and the counter counts all row activation sent to the bank. When the counter reaches a predefined value, an RFM command will be issued to refresh this bank and the counter value will be decremented accordingly. Thus, Cowles et al. apply a counting granularity of a bank, which is different from the previous solutions that count at the row-level using one counting logic per bank where the counting granularity is a single row in a bank.

\subsection{Row-Activation-Triggered Probabilistic Refresh}
In contrast to the previous category that relies on the row activation count to refresh adjacent rows deterministically, this category refreshes rows {with a fixed or adjustable probability}, indicating that they cannot enforce a rowhammer-free security invariant.

\mypara{Fixed Refreshing Probability} 
PARA~\cite{kim2014flipping} is one of the earliest probabilistic solutions to rowhammer. At every memory access, {PARA decides whether to perform an additional refresh with a low probability \texttt{P} (e.g., 0.001). If yes, it refreshes one of two rows adjacent to the activated row, and either adjacent row is selected with equal probability}. To achieve low area overhead, only a simple and compact probability generation circuit is needed to implement PARA. To enhance security, \texttt{P} can be tuned to a relatively higher value, which, however, causes more refreshes as well as energy consumption.

\mypara{Adjustable Refreshing Probability} 
As PARA {does not consider memory-access history}, it does not perform well on most benchmarks. {As such}, PRoHIT~\cite{son2017making} uses two tables labeled ``hot" and ``cold" respectively to track the access history. PRoHIT manages the tables in a probabilistic manner with some predefined values, i.e., $P_{insert}$, $P_{evict}$, and $P_{promote}$. Specifically, if a row is being activated and one of its adjacent rows is not in {either} of the two tables, it is inserted into the highest position of the cold table at a probability of $P_{insert}$. If the cold table is full, PRoHIT evicts the entry at the lowest position with the probability of $(1-P_{evict})+P_{evict}/(\#~cold\_entries$) and the other entries with a probability of $P_{evict}/(\#~cold\_entries$). If the row already exists in the cold table,  it is promoted to the lowest entry in the hot table with a probability of $(1-P_{promote})+P_{promote}/(\#~hot\_entries)$ and the other entries with a probability of $P_{promote}/(\#~hot\_entries$). If the row already exists in the hot table, it is promoted to the next higher position of the hot table. PRoHIT only refreshes the row recorded in the highest position in the hot table at every regular refresh command (issued every 7.8\,\us).

MRLoc~\cite{you2019mrloc} aims to improve the reliability of PARA while reduc{ing} the additional refreshes from PRoHIT. Similar to PRoHIT, MRLoc leverages memory-access locality by implementing a queue to track the memory-access history. Unlike PRoHIT, MRLoc leverages the queue to dynamically decide a refresh probability at every memory access. Particularly, if a row is being activated, each of its adjacent rows will be inserted into a first-in-first-out queue. If each adjacent row does not appear in the queue, it is inserted with a pre-defined refresh probability of \texttt{P}. If each adjacent row already exists in the queue (i.e., a row hit), it is still inserted into the queue and \texttt{P} for the row is adjusted based on the distance between the two inserting locations in the queue. A smaller distance means a more recently rowhammer-affected row and thus \texttt{P} will be higher. 

\subsection{{Cache Miss/Flush and DMA Access Count}}
Unlike the software detection in \autoref{sec:software_detection}, LightRoAD
~\cite{taouil2021lightroad} is a hardware scheme to detect abnormal memory-access behaviors. It resides in the MC and assigns three counters to count cache misses, cache flushes and DMA accesses, respectively. When the sum of the three counters reaches a predefined threshold, an alarm is raised. If the DMA-access count {is the major cause of the alarm}, the DMA access can be disabled for a while. 
If it is due to the cache, targeted victim data can be reallocated in the memory.

\subsection{Fabrication Process Improvement}
As the root causes of rowhammer are at the circuit-level, this category of solutions proposes new fabrication techniques~\cite{yang2016suppression,ryu2017overcoming,gautam2018improvement,gautam2019row,9025775} to mitigate rowhammer.

\mypara{Mitigating electromagnetic coupling}
One root cause is the electromagnetic coupling where the voltage fluctuation of a toggled aggressor wordline injects electromagnetic noise into its adjacent victim wordline and the noise can partially turn on the access transistor of victim cells, exacerbating the charge leakage of the capacitors in victim cells~\cite{redeker2002investigation}.
To reduce the leakage as much as possible, Yang et al.~\cite{yang2016suppression} use additional phosphorus implantation between two adjacent wordlines and junction depth optimization to form a localized shield from the electric field.
Alternatively, Gautam et al.~\cite{gautam2018improvement} introduce a low work function metal nanoparticles (MNPs) at the gate metal/gate oxide interface. The work function difference between MNP and gate-metal can create energy valleys, which effectively block the diffusion of electrons from the aggressor cells to the victim cells and thus reduce the charge leakage. However, {this solution only works for a DRAM structure that is based on the recessed channel access transistor.} To make it applicable on a DRAM structure of 3-D saddle fin recessed channel access transistor (S-RCAT), Gautam et al.~\cite{gautam2019row} replace the MNPs with metal nanowire. On top of that, they investigate the \emph{zero-failure} in S-RCAT, resulting from hammering the passing word line (PWL)~\cite{9025775}. To this end, they propose a localized introduction of a high work function PWL to suppress the impact of electric field interaction between the PWL and the charge storage node, which thus significantly mitigates the zero-failure.

\mypara{Reducing acceptor-type traps}
The other root cause is the acceptor-type trap~\cite{yang2019trap}. Specifically, in a trap-assisted charge pumping process, a trap located near the gate of an aggressor wordline captures charges and then emits them. Thus, {the emitted charges will migrate to adjacent victim wordline and cause a potential drop, leaking the charge in the capacitors of victim cells}. As a single trap may exacerbate the charge leak by a factor of 60 in a 2y-nm node, it will eventually cause bit flip if this process repeats.
To mitigate this problem, Ryu et al.~\cite{ryu2017overcoming} propose a silicon migration technique of hydrogen annealing.

\subsection{Summary of Hardware Defenses}
All existing hardware defenses except row-activation-count based defenses cannot eliminate rowhammer. Specifically, the data integrity check based solutions either cannot detect and correct all induced bit flips~\cite{kim2019effective,wang2019reinforce,ryan2009channel} or have only detection capability~\cite{gomez2016dram,vig2018rapid}. The row-activation-triggered based probabilistic refresh approaches are intrinsically flawed with security. Particularly, PARA~\cite{kim2014flipping} incur{s} a high performance and energy cost when mitigating rowhammer to a larger extent. Though PRoHit~\cite{son2017making} and MRLoc~\cite{you2019mrloc} have optimized PARA significantly, they are still vulnerable to specific adversarial memory access patterns~\cite{park2020graphene}. The fabrication-process-improvement based methods only address the root causes of rowhammer to some extent.

While row-activation-count-based defenses aim to provide a rowhammer-free security guarantee, most of them hardly scale to an increasingly low MAC. As a DRAM chip becomes denser, the MAC decreases considerably and thus the blast radius increases~\cite{kim2020revisiting}. Taking a lower MAC and higher blast radius into consideration, such defenses inevitably introduce more expensive refreshes or throttling caused by benign applications.
\end{document}